\documentclass[10pt]{article}
\usepackage{float}
\usepackage[symbol]{footmisc}
\usepackage[superscript,sort]{cite}
\usepackage{array}
\usepackage[normalem]{ulem} 
\usepackage[colorlinks=true,linkcolor=black,citecolor=black,urlcolor=black,bookmarks=true,breaklinks=true]{hyperref}
\usepackage[usenames]{color}
\usepackage{amsmath,amssymb}

\usepackage{amsthm,amscd,amsxtra,amsfonts,amsmath,amssymb,multirow}
\usepackage{wrapfig}
\usepackage[footnotesize]{caption}
\usepackage{subcaption}
\usepackage[tiny,compact]{titlesec}
\usepackage{threeparttable} 
\usepackage{helvet}

\usepackage{xcolor,colortbl}
\usepackage{tikz}
\usetikzlibrary{shapes,arrows}
\usepackage[T1]{fontenc}
\titlespacing*{\section}{0pt}{*0}{*0}
\titlespacing*{\subsection}{0pt}{*0}{*0}
\titlespacing*{\subsubsection}{0pt}{*0}{*0}
\titlespacing{\paragraph}{0pt}{*0}{*1}

\definecolor{MyPurple}{rgb}{1,0,1}

\newcommand{\beq}[1]{\begin{equation} \label{#1}}
\newcommand{\eeq}{\end{equation}}
\newcommand{\barray}{\begin{array}{ll}}
\newcommand{\earray}{\end{array}}

\begin{document}
\pagenumbering{roman}

\clearpage \pagebreak \setcounter{page}{1}
\renewcommand{\thepage}{{\arabic{page}}}

\title{Generalized flexibility-rigidity index
}

\author{
Duc Duy Nguyen$^1$,
Kelin Xia$^1$  and
Guo-Wei Wei$^{1,2,3}$ \footnote{ Address correspondences  to Guo-Wei Wei. E-mail:wei@math.msu.edu}\\
$^1$ Department of Mathematics \\
Michigan State University, MI 48824, USA\\
$^2$  Department of Biochemistry and Molecular Biology\\
Michigan State University, MI 48824, USA \\
$^3$ Department of Electrical and Computer Engineering \\
Michigan State University, MI 48824, USA \\
}

\date{\today}
\maketitle

\begin{abstract}

Flexibility-rigidity index (FRI) has been developed as a robust, accurate  and efficient method for macromolecular thermal fluctuation analysis and B-factor prediction. The performance of FRI depends on its formulations of rigidity index and flexibility index. In this work, we introduce alternative rigidity and flexibility  formulations. The structure of the classic Gaussian surface is utilized to construct a new type of rigidity index, which leads to a new class of rigidity densities with the classic Gaussian surface as a special case. Additionally,  we introduce a new type of flexibility index based on the domain indicator property of normalized rigidity density. These generalized FRI (gFRI) methods  have  been extensively validated by the B-factor predictions of 364 proteins. Significantly outperforming the classic  Gaussian network model (GNM),  gFRI is a new generation of methodologies for accurate, robust and efficient analysis of protein flexibility and fluctuation. Finally, gFRI based molecular surface generation and flexibility visualization  are demonstrated.

\end{abstract}
\maketitle



In  living organisms, proteins carry out a vast variety of basic functions, such as structure support,   catalyzing chemical reactions,  and allosteric regulation, through synergistic interactions or  correlations.  Protein functions and interactions are determined by protein structure and flexibility \cite{Frauenfelder:1991}. The importance of protein structure needs no introduction, while the importance of protein
flexibility is often overlooked. Protein flexibility is an intrinsic property of proteins  and  can be measured by experimental means, including X-ray crystallography,  nuclear magnetic resonance (NMR) and single-molecule force spectroscopy, e.g., magnetic tweezer, optical trapping and atomic force microscopy \cite{Dudko:2006}.  Flexibility analysis offers a unique channel for theoretical modeling to meet with experimental observations. A variety of theoretical methods, such as  normal mode analysis (NMA)  \cite{Go:1983,Tasumi:1982,Brooks:1983,Levitt:1985,JMa:2005}, graph theory \cite{Jacobs:2001},  { rotation translation blocks (RTB) method \cite{tama:2000,Demerdash:2012}, and elastic network model (ENM) \cite{Bahar:1997,Bahar:1998,Atilgan:2001,Hinsen:1998,Tama:2001,LiGH:2002},  including    Gaussian network model (GNM)   \cite{Bahar:1997,Bahar:1998}  and anisotropic network model (ANM) \cite{Atilgan:2001}, have been proposed. Among them, GNM is often favored due to its  accuracy and efficiency \cite{LWYang:2008}. These time-independent methods have been widely used not only for protein fluctuation analysis, but also for entropy estimation. However, they typically suffer from two major drawbacks: 1) ${\cal O}(N^3)$ scaling in computational complexity with $N$ being the number of elements in the involved matrix and 2) insufficient accuracy in protein B-factor predictions. The above scaling in computational complexity is due to the matrix diagonalization and makes large biomolecules inaccessible to aforementioned methods. Recently, Park {\it et al.} have shown that for three sets of structures of small-sized, medium-sized and large-sized, the mean correlation coefficients (MCCs) for  NMA and GNM B-factor predictions are respectively  below 0.5 and 0.6 \cite{JKPark:2013}. These researchers  found that   both NMA and GNM fail to work for many structures and deliver  negative correlation coefficients   \cite{JKPark:2013}. These problems call for the  development of  accurate, efficient and reliable  approaches for the flexibility analysis and entropy calculation of macromolecules.

One strategy to tackle  the above-mentioned challenges is to develop matrix-diagonalization-free methods for flexibility analysis. To this end,
we have introduced   molecular nonlinear dynamics \cite{KLXia:2014b},  stochastic dynamics \cite{KLXia:2013f} and flexibility-rigidity index  (FRI)  \cite{KLXia:2013d,Opron:2014}. {\color{black}Our approaches make use of protein network connectivity and centrality to describe protein flexibility and rigidity}.  In our FRI method, we assume that protein interactions, including those with its environment,  fully determine {\color{black}  its structure in the given  environment.  In contrast,  protein  flexibility and rigidity } are fully determined by the structure of the protein and its environment. Therefore, to analyze protein flexibility and rigidity, it is unnecessary to resort to the protein interaction Hamiltonian whenever an accurate  protein structure is already available. As a result,  FRI bypasses the  ${\cal O}(N^3)$ matrix diagonalization.
Our earlier FRI  \cite{KLXia:2013d} has the computational complexity of ${\cal O}(N^2)$ and our fast FRI (fFRI) \cite{Opron:2014} based on a cell lists algorithm \cite{Allen:1987} is of ${\cal O}(N)$. Anisotropic FRI (aFRI) \cite{Opron:2014} and multiscale FRI (mFRI) \cite{Opron:2015a} have also been proposed. FRI correlation kernels are utilized to develop generalized  GNM (gGNM) and generalized ANM (gANM) methods as well as multiscale GNM (mGNM) and multiscale ANM (mANM) methods \cite{KLXia:2015f}, which significantly improves their accuracy. In the past two years, we have extensively validated  FRI,  fFRI, aFRI and mFRI  by a set of 364 proteins for accuracy, reliability and efficiency. Our mFRI is about 20\%  more accurate than   GNM on the 364 protein test set \cite{Opron:2015a}.  Our fFRI is orders of magnitude faster than GNM on a set of 44  proteins, including one of the largest proteins in the Protein Data Bank (PDB), namely, an HIV virus capsid (1E6J)  having 313,236 residues. Our fFRI completes the B-factor prediction of the HIV capsid within  30 seconds on a single-core  processor, which would take  GNM more than 120 years  to accomplish had the computer memory not been a problem \cite{Opron:2014}.

Although various forms of FRI correlation kernels have been introduced, the general mathematical structure of FRI  has not been studied. For example, only one rigidity formula and one flexibility formula  were proposed\cite{KLXia:2013d,Opron:2014}. It is interesting to know whether there  exists alternative  FRI formulations. If so, how do they perform against other existing methods in experimental B-factor predictions?  The objective  of the present work is to shed lights on these issues. Motivated by the structure of  the popular Gaussian surface  \cite{Zap,Grant:2007, LLi:2013, LinWang:2015}, we  propose an alternative  rigidity index and a new rigidity density. The latter systematically extends the Gaussian surface to surface densities equipped with a wide variety of FRI correlation kernels. Additionally, we propose normalized rigidity index and normalized rigidity density. The latter behaves like a protein domain indicator \cite{Wei:2009}, which inspires us to introduce a new form of flexibility index and flexibility function. These generalized FRI (gFRI) formulations are extensively validated against experimental data and their performances are systematically compared with a number of other methods.  {\color{black} In addition, the new form of flexibility index has been incorporated into aFRI to predict the amplitudes and the directions of atomic fluctuation.}



In a molecule with $N$   atoms, we denote $\mathbf{r}_j \in \mathbb{R}^3$ the position of $j$th   atom, and $\|\mathbf{r}_i -\mathbf{r}_j\|$ the Euclidean distance between $i$th   and $j$th  atom. An atomic rigidity  index is defined as \cite{KLXia:2013d,Opron:2014}
\begin{align}\label{rigidity1}
\mu^1_i=\sum_{\substack{j=1}}^{N} w_{ j} \Phi\left(\|\mathbf{r}_i -\mathbf{r}_j\|;\eta_{ j}\right),
\end{align}
where $w_{j}$ are particle-type related weights that can be set to $w_{j}=1$ for the present work, $\eta_{j}$ are characteristic distances   and $\Phi$ is a correlation kernel that satisfies the following admissibility conditions
\begin{align}\label{eq:admiss}
\Phi \left(\|\mathbf{r}_i - \mathbf{r}_i\|;\eta_{j}\|\right)&=1, \quad{\rm as} \quad  \|\mathbf{r}_i -\mathbf{r}_j\| \rightarrow 0, \\
\Phi \left(\|\mathbf{r}_i - \mathbf{r}_j\|;\eta_{j}\|\right)&=0, \quad {\rm as} \quad  \|\mathbf{r}_i -\mathbf{r}_j\| \rightarrow \infty.
\end{align}
Monotonically decaying  radial basis functions are all admissible. Commonly used FRI correlation kernels include  generalized exponential functions
\begin{align}
\Phi\left(\|\mathbf{r}_i -\mathbf{r}_j\|;\eta_{ j}\|\right)=e^{-\left(\|\mathbf{r}_i -\mathbf{r}_j\|/\eta_{ j}\right)^\kappa}, \quad \kappa>0;
\end{align}
and generalized Lorentz functions
\begin{align}
\Phi\left(\|\mathbf{r}_i -\mathbf{r}_j\|;\eta_{ j}\right)=\frac{1}{1+\left(\|\mathbf{r}_i -\mathbf{r}_j\|/\eta_{ j}\right)^\nu},\quad \nu>0.
\end{align}
Many other functions, such as delta sequences of the positive type discussed in an earlier work \cite{GWei:2000} can be employed as well.

The rigidity index in Eq. (\ref{rigidity1}) was extended into a continuous rigidity density \cite{KLXia:2013d,Opron:2014}
 \begin{align}\label{fri_surface}
	\mu^1(\mathbf{r})=\sum_{\substack{j=1}}^{N} w_{j} \Phi\left(\|\mathbf{r}-\mathbf{r}_j\|;\eta_{j}\right).
\end{align}
 It has been shown that rigidity density (\ref{fri_surface}) serves as an excellent representation of molecular surfaces \cite{KLXia:2015e}. This connection motivates us to generalize the Gaussian  surface    \cite{Zap,Grant:2007, LLi:2013, LinWang:2015} to a new class of surface densities equipped with a wide variety of FRI correlation kernels  ($\Phi\left(\|\mathbf{r} -\mathbf{r}_j\|;\eta_{ j}\right)$)
\begin{align}\label{rigidity2}
\mu^2(\mathbf{r}) =1-\prod_{\substack{j=1 \\ {\bf r}\neq {\bf r}_j}}^{n}\left[1-w_{ j}\Phi\left(\|\mathbf{r} -\mathbf{r}_j\|;\eta_{ j}\right)\right].
\end{align}
Since both rigidity densities $\mu^\alpha(\mathbf{r}), ~\alpha=1,2$  represent molecular density at position $\mathbf{r}$,  it is convenient to normalize these densities by their maximal values
\begin{align}\label{normalization}
	\bar{\mu}^{\alpha}(\mathbf{r})=\frac{\mu^{\alpha}(\mathbf{r})}{\max\limits_{\mathbf{r}\in \mathbb{R}^3} \mu^\alpha(\mathbf{r})}, \quad \alpha =1,2.
\end{align}
In this form, the behaviors of two types of rigidity based  molecular surfaces can be easily compared. Additionally,  normalized rigidity densities in Eq. (\ref{normalization})  can be used as solute domain indicators in implicit solvent models \cite{Wei:2009}. Obviously, $\max\mu^\alpha(\mathbf{r})$ occurs at an atomic position.  Therefore, we can define normalized atomic  rigidity indexes
\begin{align}\label{normalizedR}
\bar{\mu}^\alpha_i=\bar{\mu}^\alpha(\mathbf{r}_i), \quad  \alpha =1,2.
\end{align}

With   atomic rigidity indexes, $\bar{\mu}^\alpha_i$, we denote the flexibility indexes proposed in our earlier work \cite{KLXia:2013d,Opron:2014} as   atomic flexibility indexes of  type I  \cite{KLXia:2013d,Opron:2014}
\begin{align}\label{flexibility1}
f^{\alpha 1}_i=\frac{1}{\bar{\mu}^\alpha_i},   \quad \forall i=1,2,\dots,N; \quad \alpha=1,2.
\end{align}
The definition of atomic flexibility indexes is not unique. One of the present objectives is to explore other  forms of atomic flexibility indexes.   Since  the normalized atomic rigidity density  can be interpreted as a solute domain indicator, then $1-\bar{\mu}^\alpha(\mathbf{r})$ can be regarded as a solvent domain indicator \cite{Wei:2009,ZhanChen:2010a}. This motivates us to propose a new form of atomic  flexibility indexes
\begin{align}\label{flexibility2}
f^{\alpha 2}_i=1-\bar{\mu}^\alpha_i, \quad   \forall i=1,2,\dots,N; \quad \alpha=1,2.
\end{align}
We denote $f^{\alpha 2}_i$ as  atomic flexibility indexes of  type II.

\begin{table}[!htb]
		\centering
\begin{threeparttable}[b]
	\caption{Mean correlation coefficients (MCCs) for protein B-factor predictions.}
	\label{tab.results}
	\begin{tabular}{lllll}
		\hline
		\multicolumn{1}{c}{Method} & Exponential kernels & MCC & Lorentz kernels & MCC\\
		\hline
		gFRI$^{11}$	      &  $\kappa=1.0, \eta=3.0$ \AA & 0.625 & $\nu=3.0, \eta=3.0$ \AA  & 0.628 \\
		gFRI$^{12}$       &  $\kappa=1.0, \eta=4.0$ \AA & 0.607 & $\nu=2.5, \eta=1.0$ \AA  & 0.613 \\
		gFRI$^{21}$       &  $\kappa=1.0, \eta=3.0$ \AA & 0.604 & $\nu=2.5, \eta=1.0$ \AA  & 0.626 \\
		gFRI$^{22}$       &  $\kappa=1.0, \eta=3.0$ \AA & 0.621 & $\nu=2.5, \eta=2.0$ \AA  & 0.627 \\
		FRI\tnote{$a$}    &  $\kappa=1.0,	\eta=3.0$ \AA & 0.623 & $\nu=3.0, \eta=3.0$ \AA  & 0.626 \\
		gGNM\tnote{$b$}		&  $\kappa=1.0,	\eta=3.0$	\AA & 0.608 & $\nu=3.0, \eta=0.5\text{~\AA}$ & 0.622  \\
		gANM\tnote{$c$}		&  $\kappa=2.0, \eta=11.0\text{~\AA}$	& 0.518 & Not available \\
		GNM\tnote{$d$}		&   Not applicable 	& 0.565 & ~ \\
		\hline
	\end{tabular}
	\begin{tablenotes}
			\item [$a$] Results   averaged over 365 proteins from  Ref. \cite{Opron:2014}.
			\item [$b$] Results   averaged over 362 proteins from  Ref. \cite{KLXia:2015f}.
			\item [$c$] Results   averaged over 300 proteins from  Ref. \cite{KLXia:2015f}.
			\item [$d$] Results  obtained with   cutoff distance 7\AA~ averaged over 365 proteins from  Ref. \cite{Opron:2014}.
	\end{tablenotes}
\end{threeparttable}
\end{table}

\begin{figure}[!tb]
	\centering
	\begin{subfigure}[!tb]{0.6\textwidth}
		\includegraphics[width=0.48\textwidth]{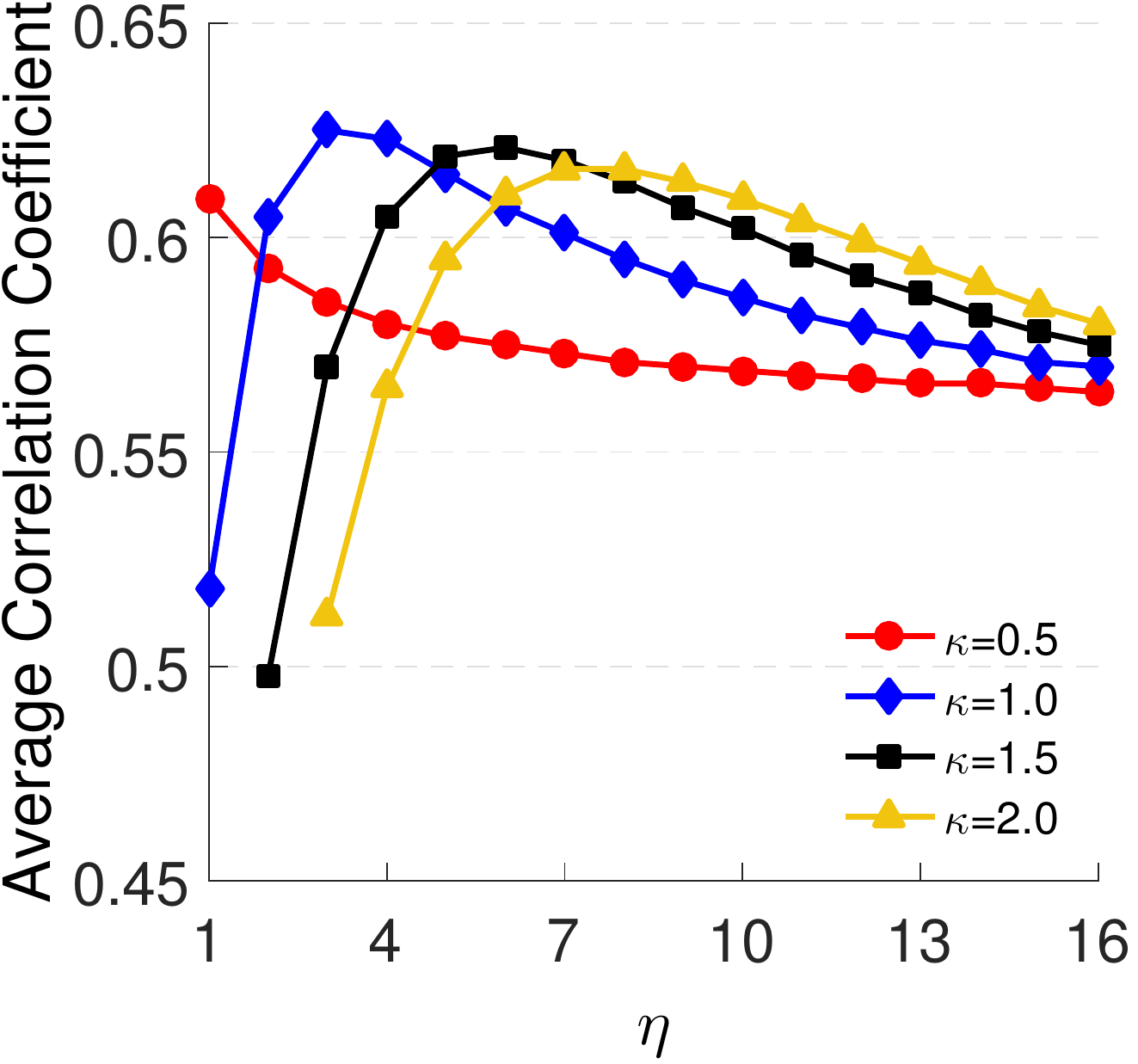}\quad
		\includegraphics[width=0.48\textwidth]{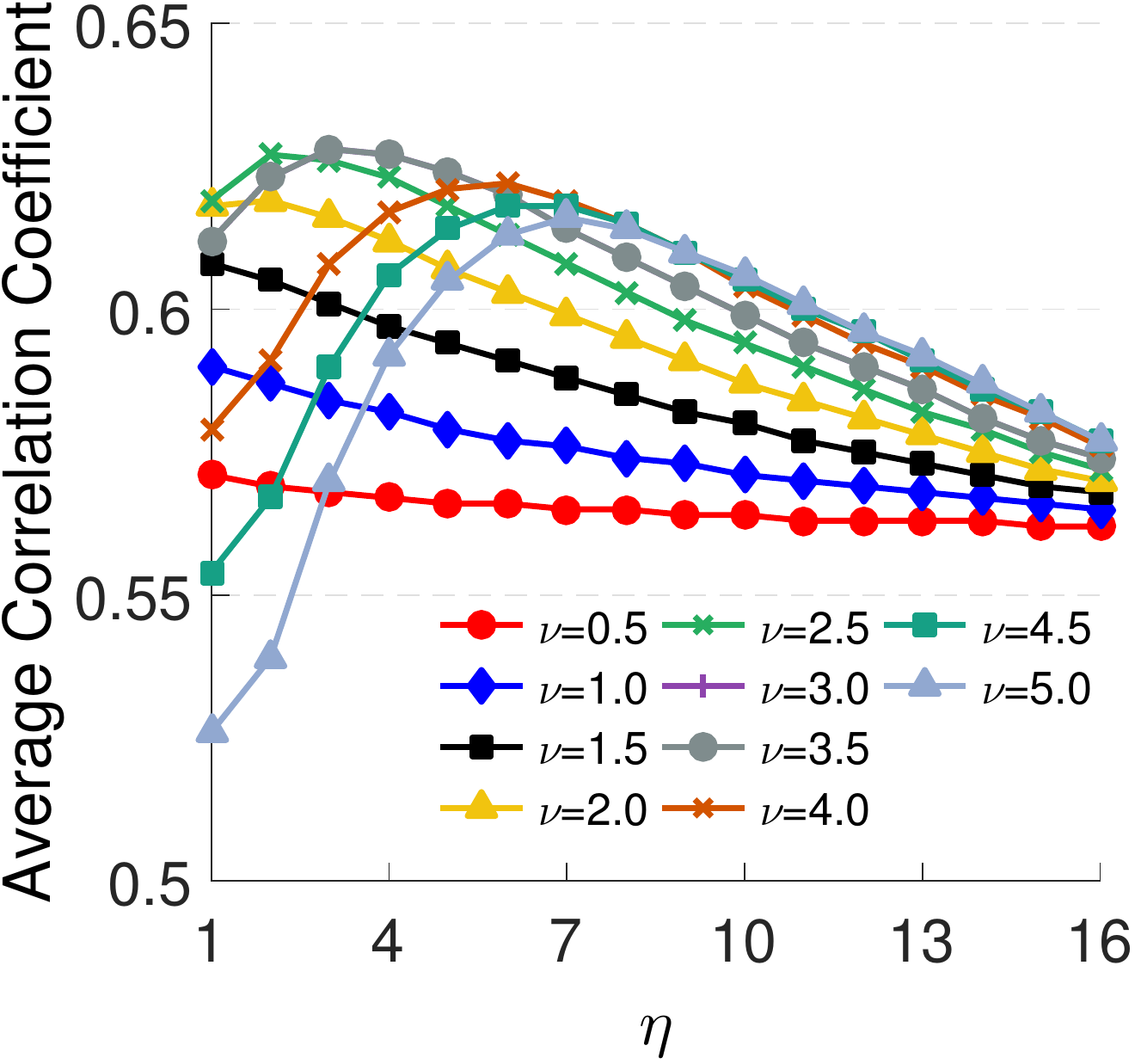}
		\caption{gFRI$^{11}$}
	\end{subfigure}
	\begin{subfigure}[!tb]{0.6\textwidth}
		\includegraphics[width=0.48\textwidth]{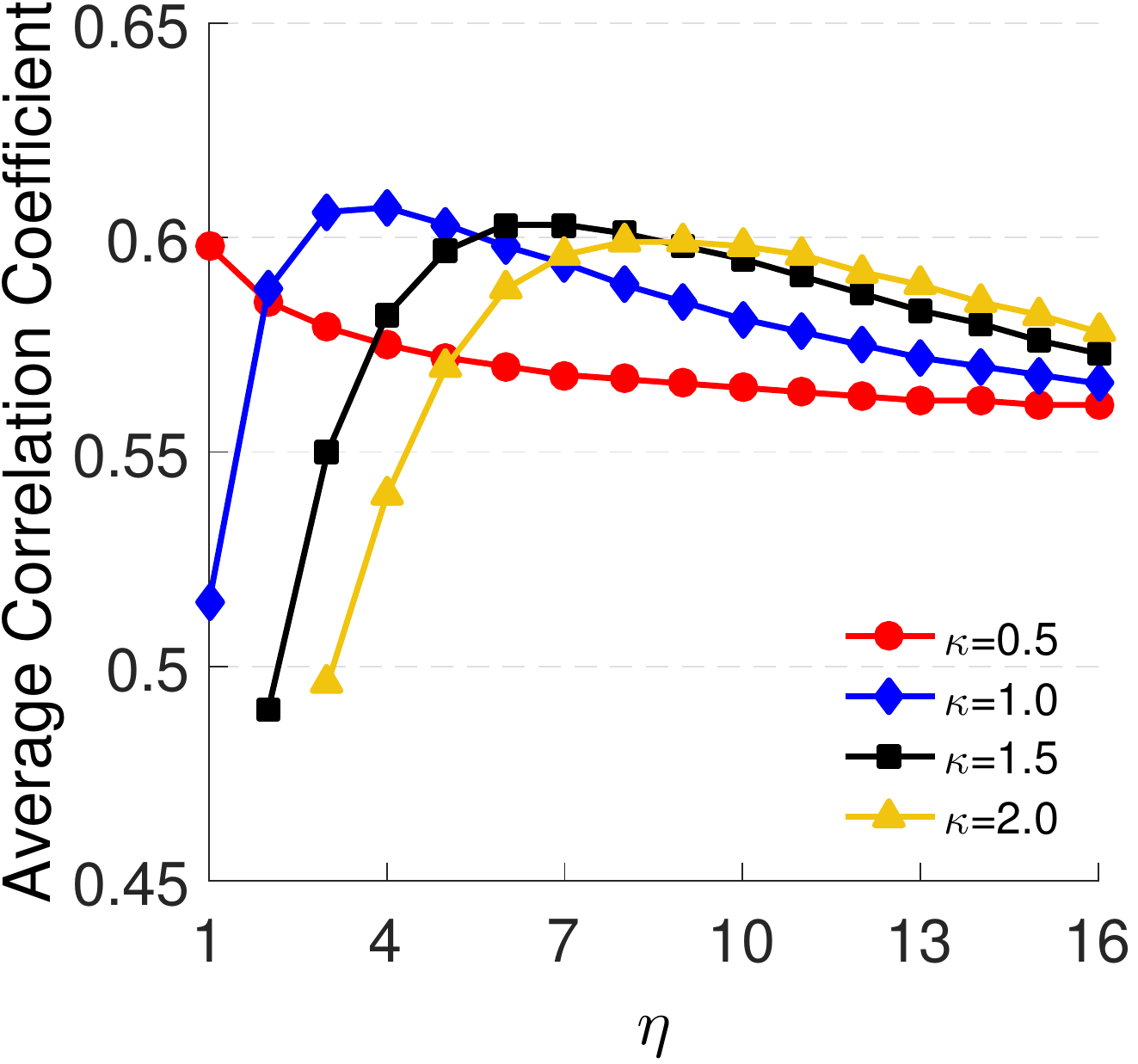}\quad
		\includegraphics[width=0.48\textwidth]{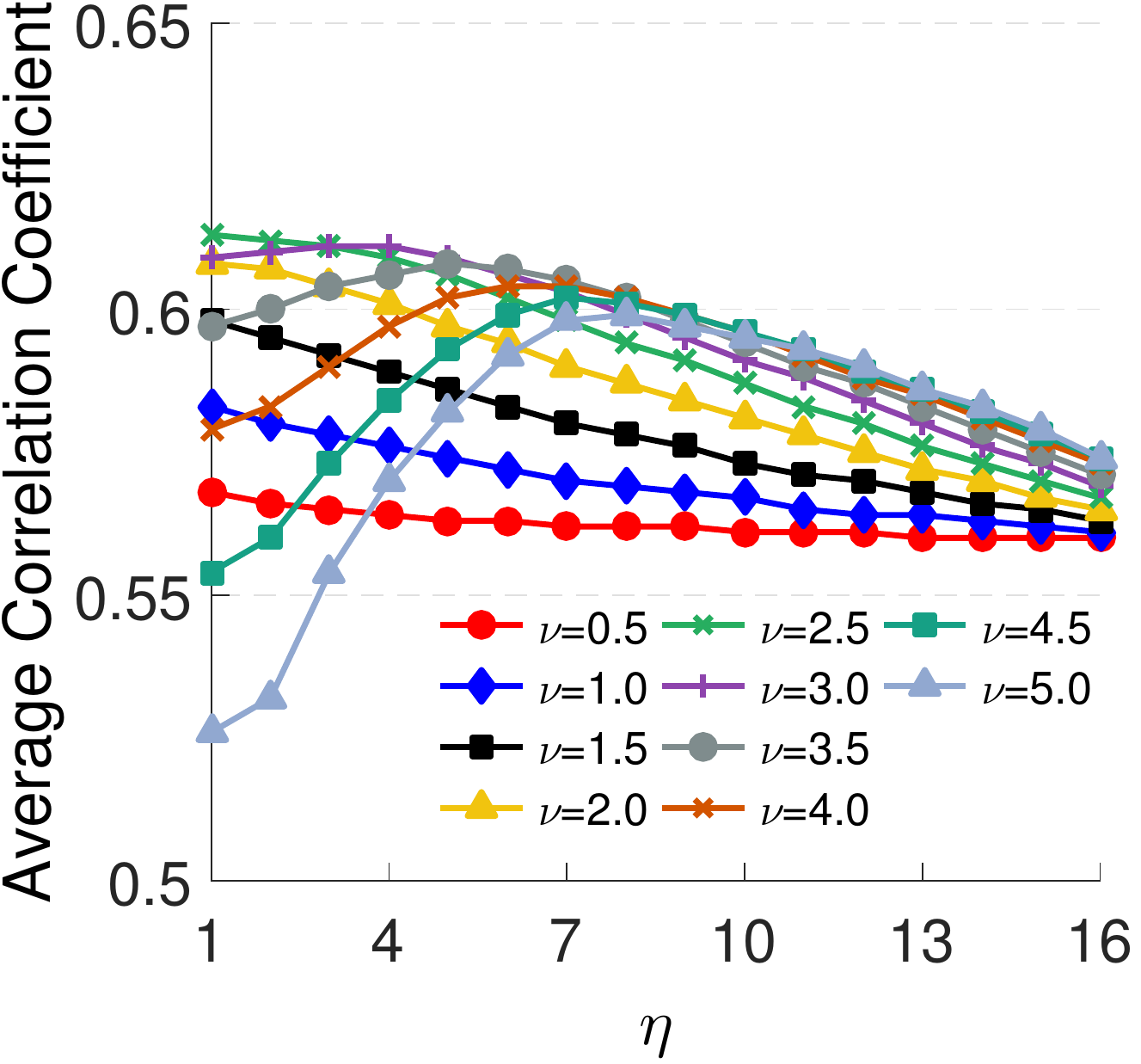}
		\caption{gFRI$^{12}$}
	\end{subfigure}
	\begin{subfigure}[!tb]{0.6\textwidth}
		\includegraphics[width=0.48\textwidth]{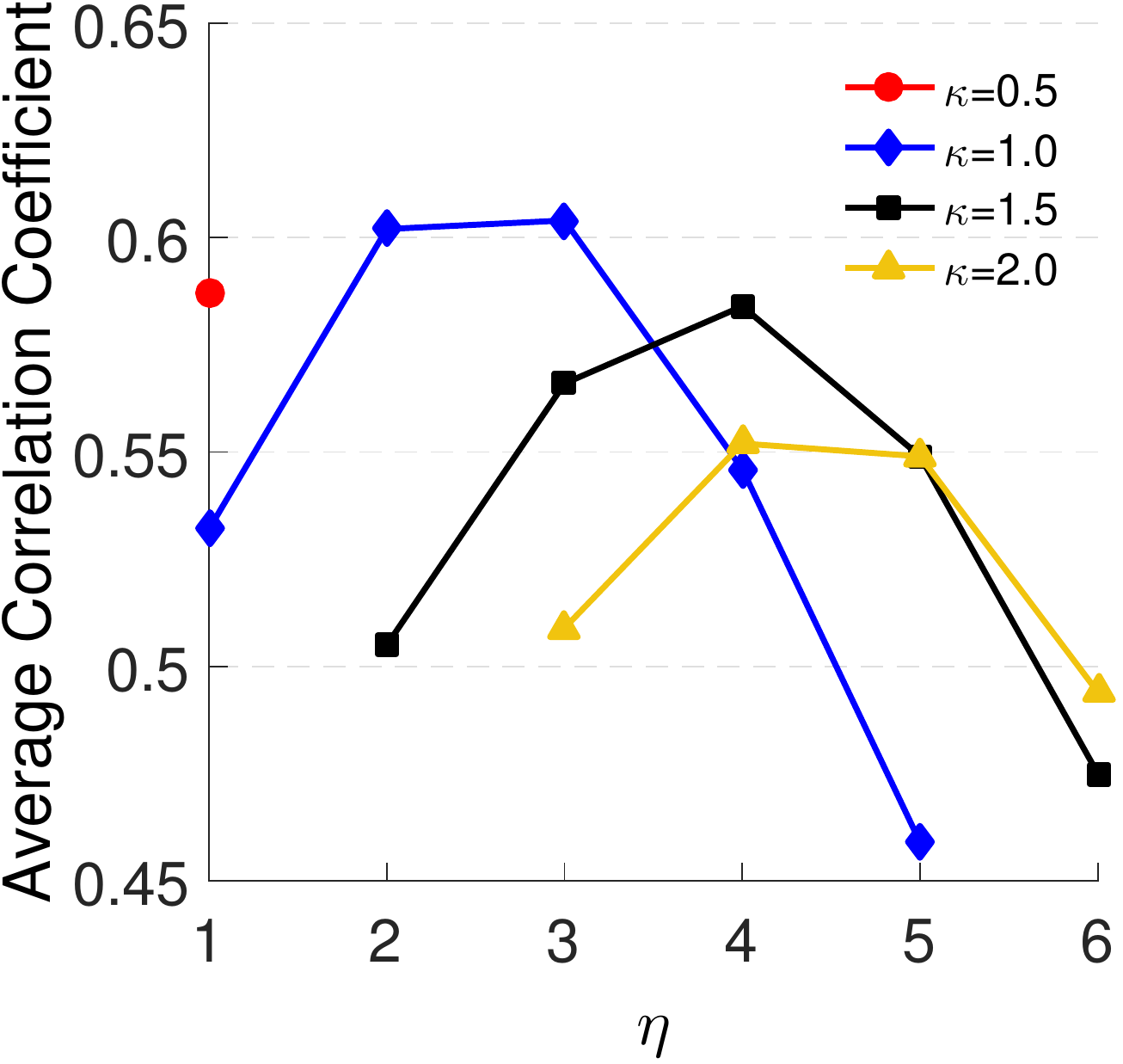}\quad
		\includegraphics[width=0.48\textwidth]{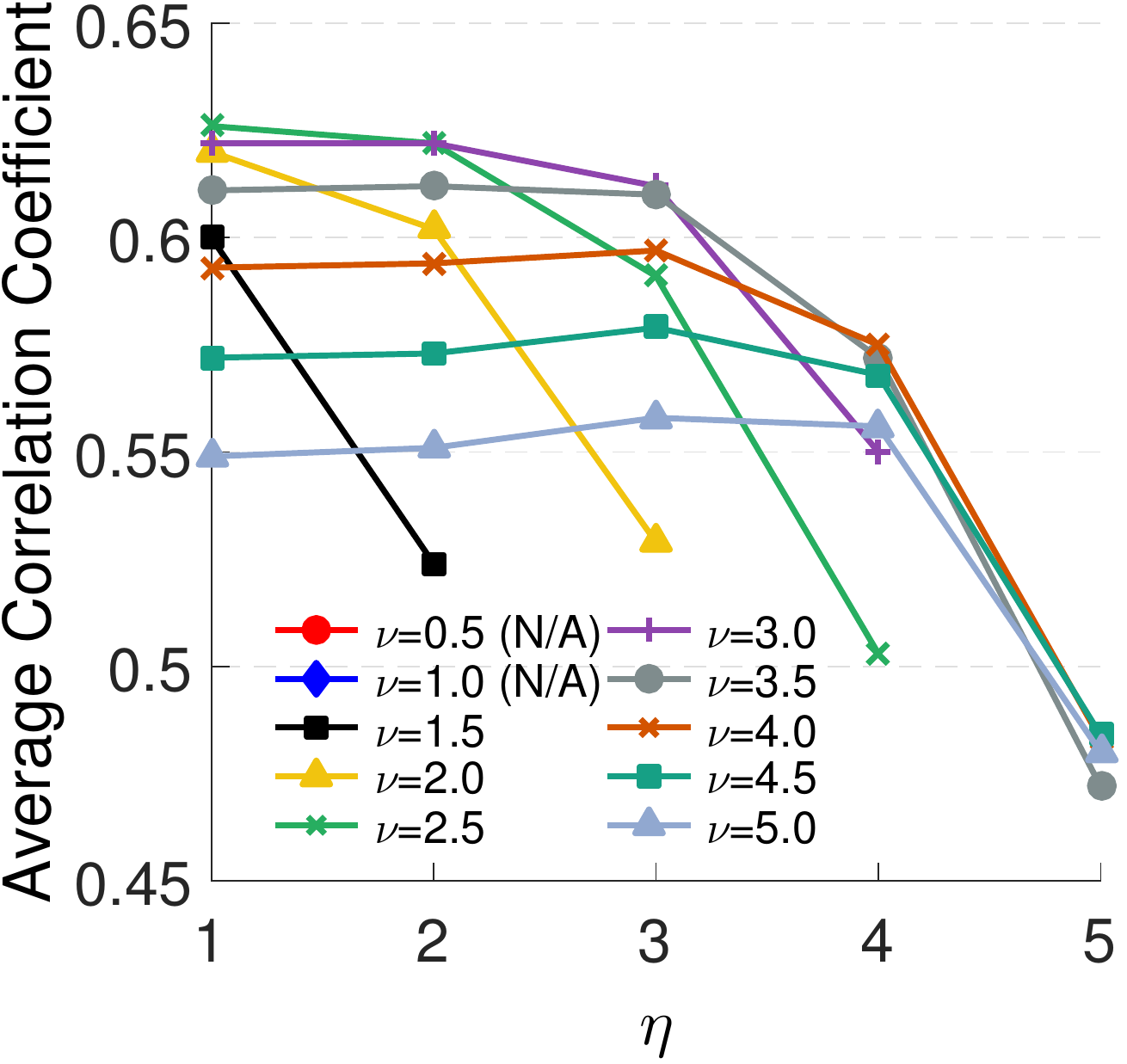}
		\caption{gFRI$^{21}$}
	\end{subfigure}
	\begin{subfigure}[!tb]{0.6\textwidth}
		\includegraphics[width=0.48\textwidth]{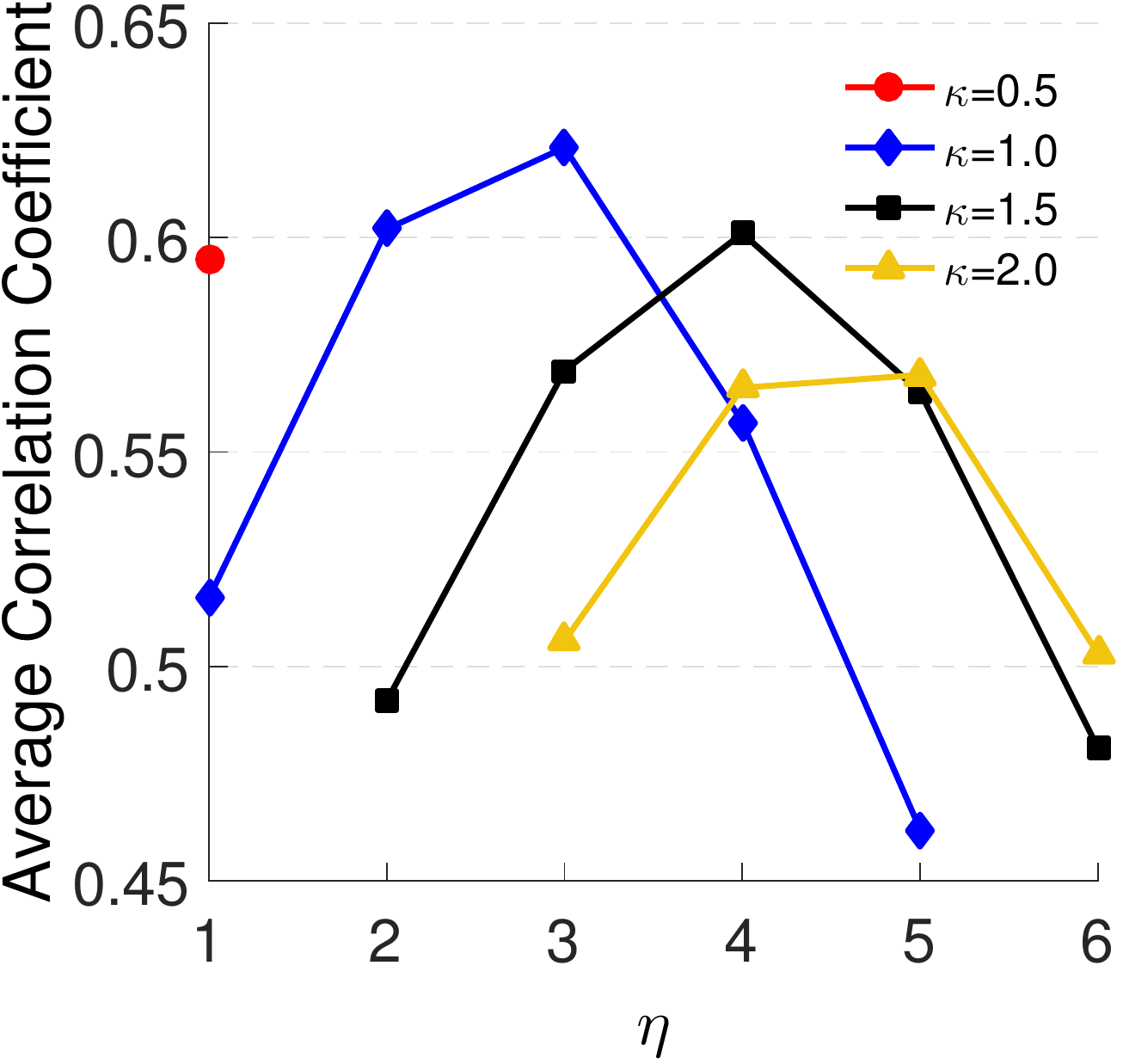}\quad
		\includegraphics[width=0.48\textwidth]{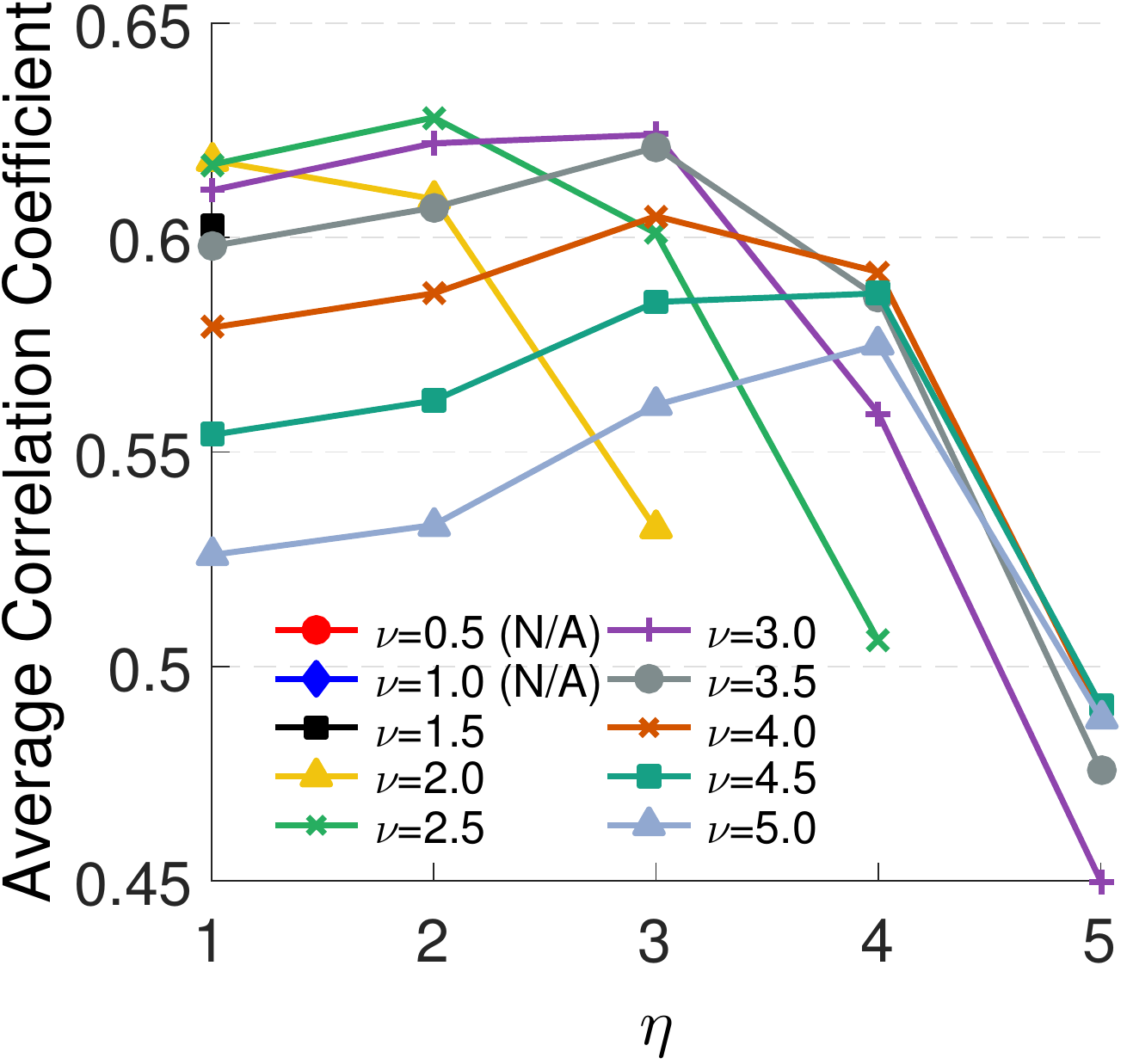}
		\caption{gFRI$^{22}$}
	\end{subfigure}
	\caption{Parameter testing of gFRI methods for exponential (left chart) and Lorentz (right
		chart) functions. Mean correlation coefficients (MCCs) of B-factor predictions of 364
		proteins are plot against choice of $\eta$ for a range of values for $\kappa$ or $\nu$.
		Note that results with MCCs less 0.45 are not shown.}
	\label{fig.cc}
\end{figure}

\begin{figure}[!htb]
	\centering
	\begin{subfigure}[!htb]{0.8\textwidth}
		\includegraphics[width=0.48\textwidth]{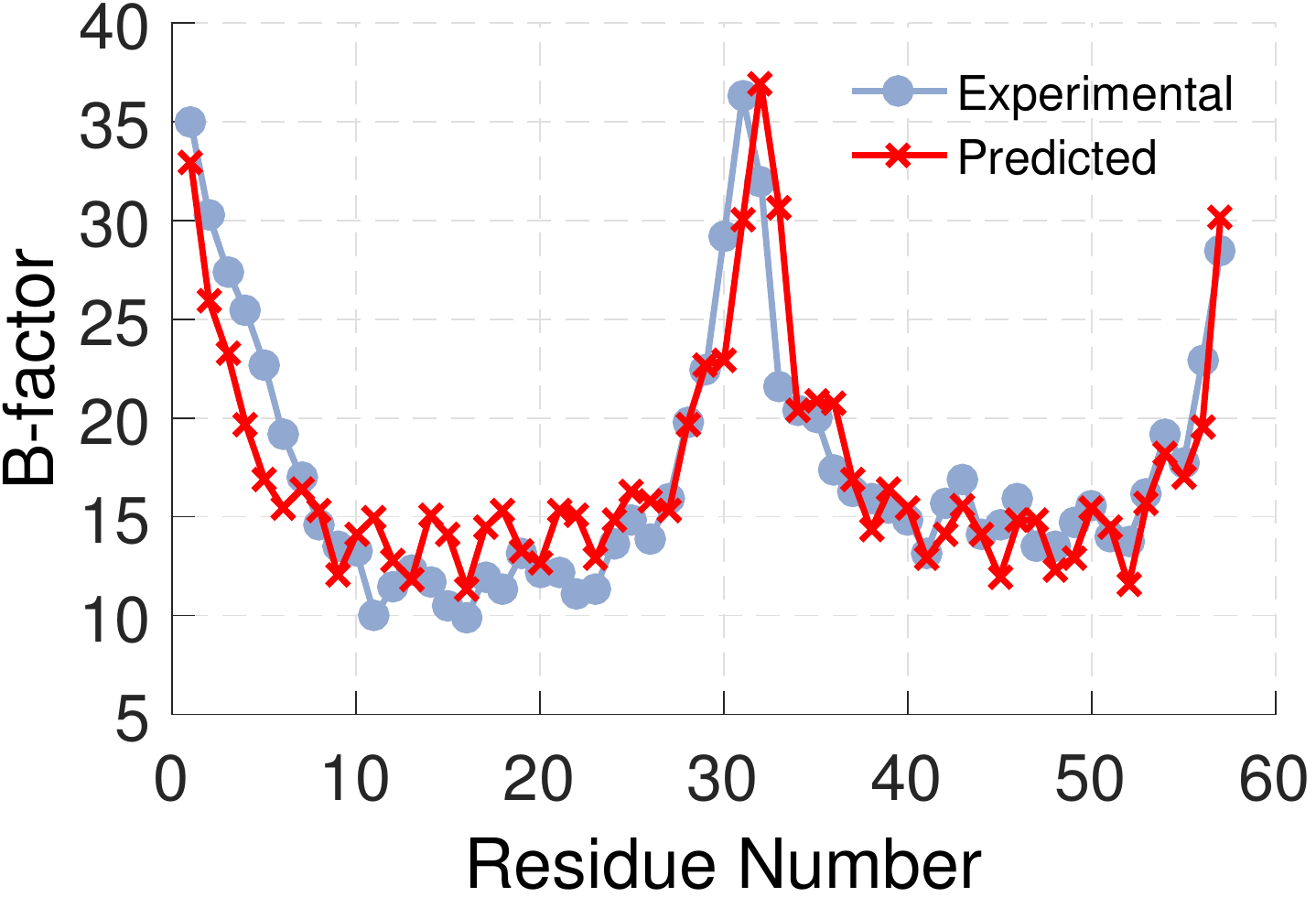}\quad
		\includegraphics[width=0.48\textwidth]{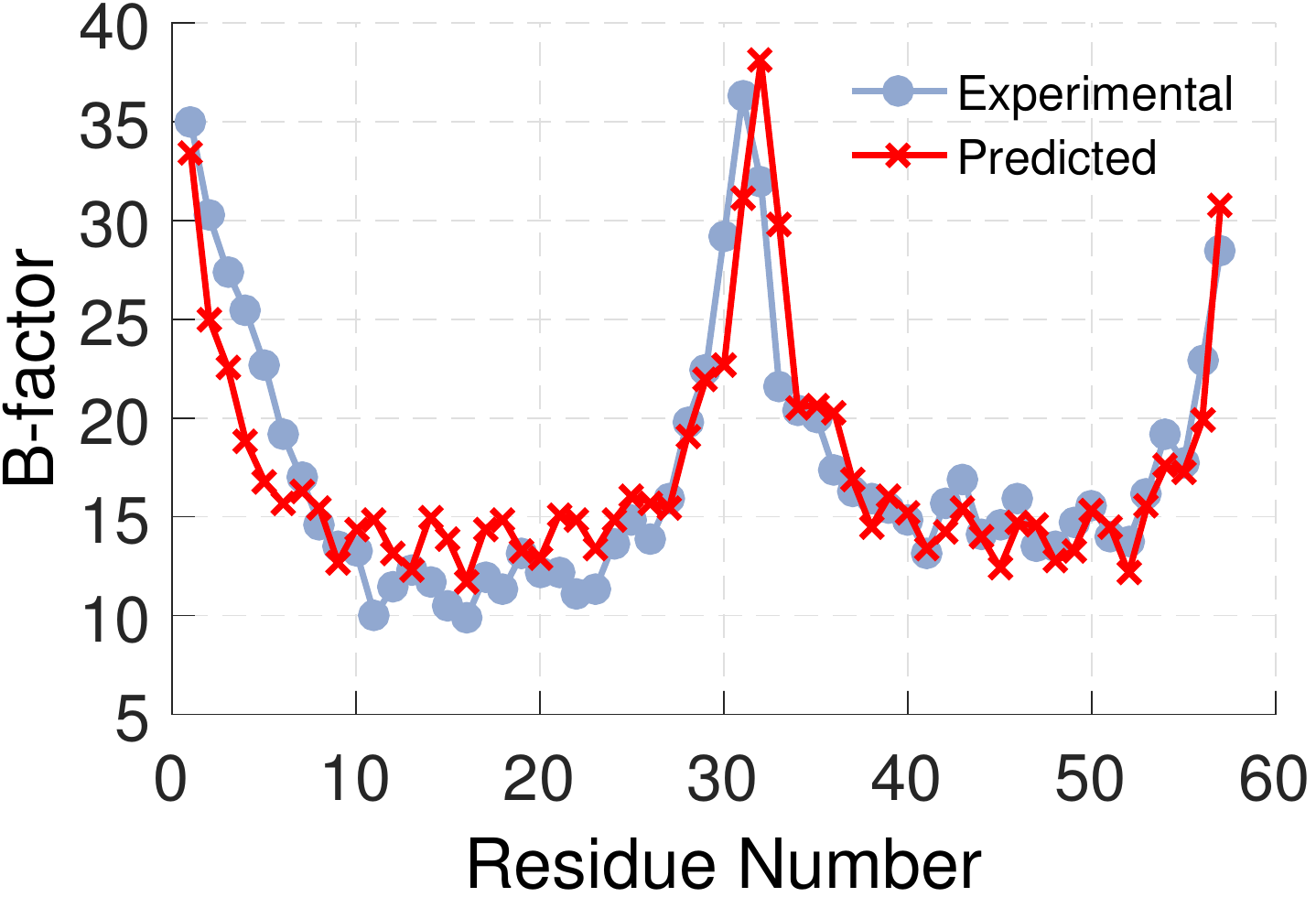}
		\caption{gFRI$^{11}$}
	\end{subfigure}
	\begin{subfigure}[!tb]{0.8\textwidth}
		\includegraphics[width=0.48\textwidth]{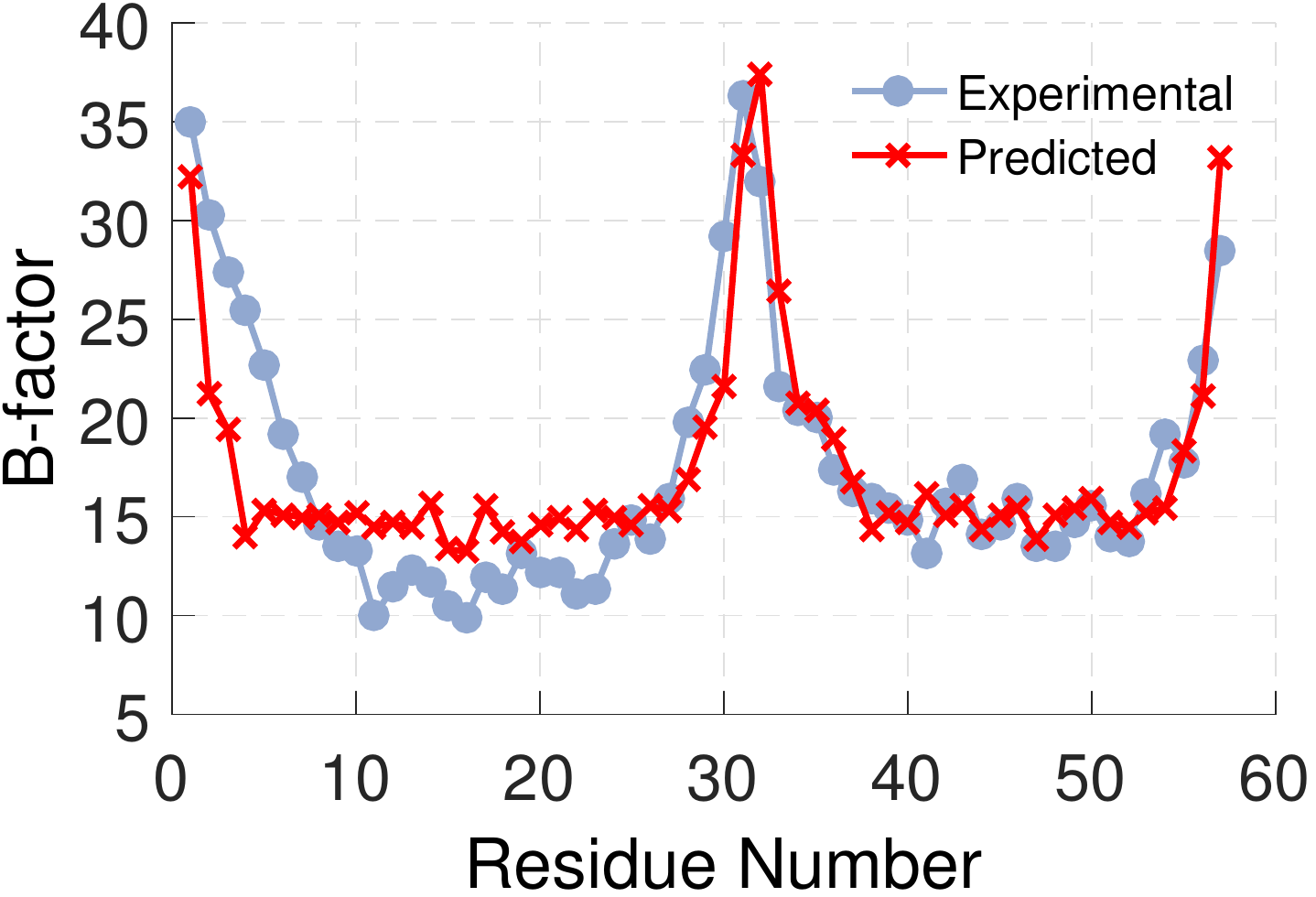}\quad
		\includegraphics[width=0.48\textwidth]{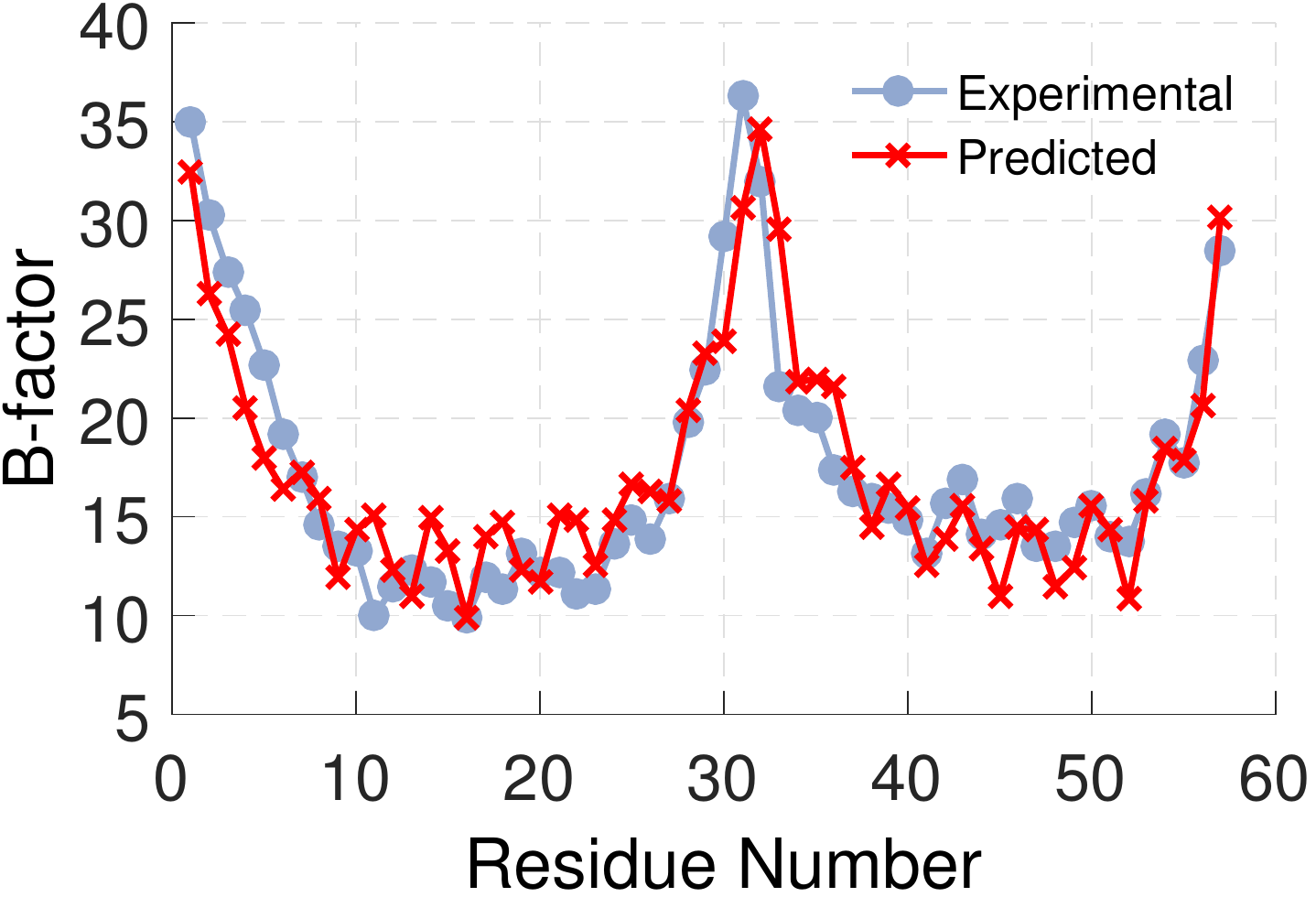}
		\caption{gFRI$^{12}$}
	\end{subfigure}
	\begin{subfigure}[!tb]{0.8\textwidth}
		\includegraphics[width=0.5\textwidth]{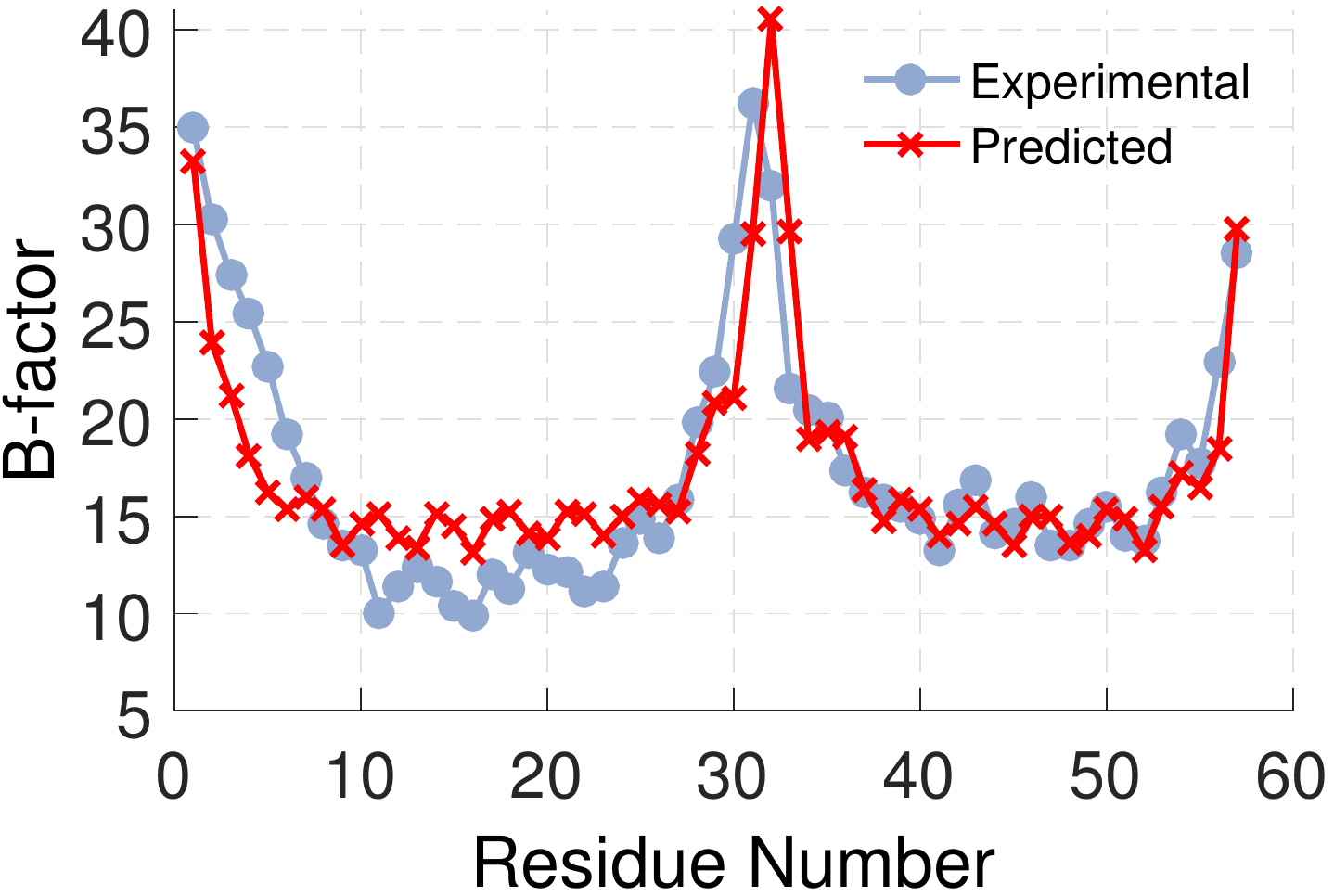}\quad
		\includegraphics[width=0.5\textwidth]{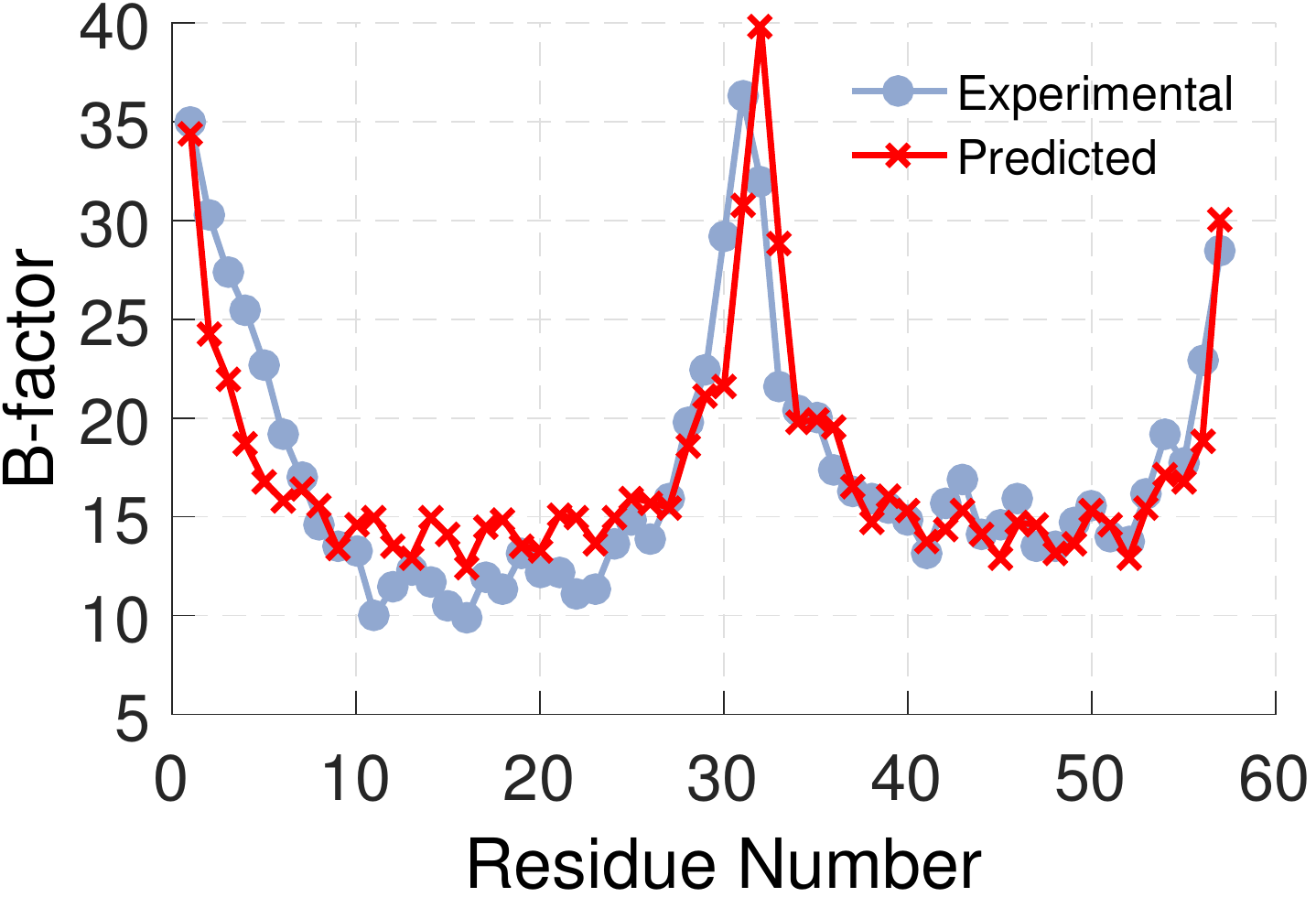}
		\caption{gFRI$^{21}$}
	\end{subfigure}
	\begin{subfigure}[!tb]{0.8\textwidth}
		\includegraphics[width=0.5\textwidth]{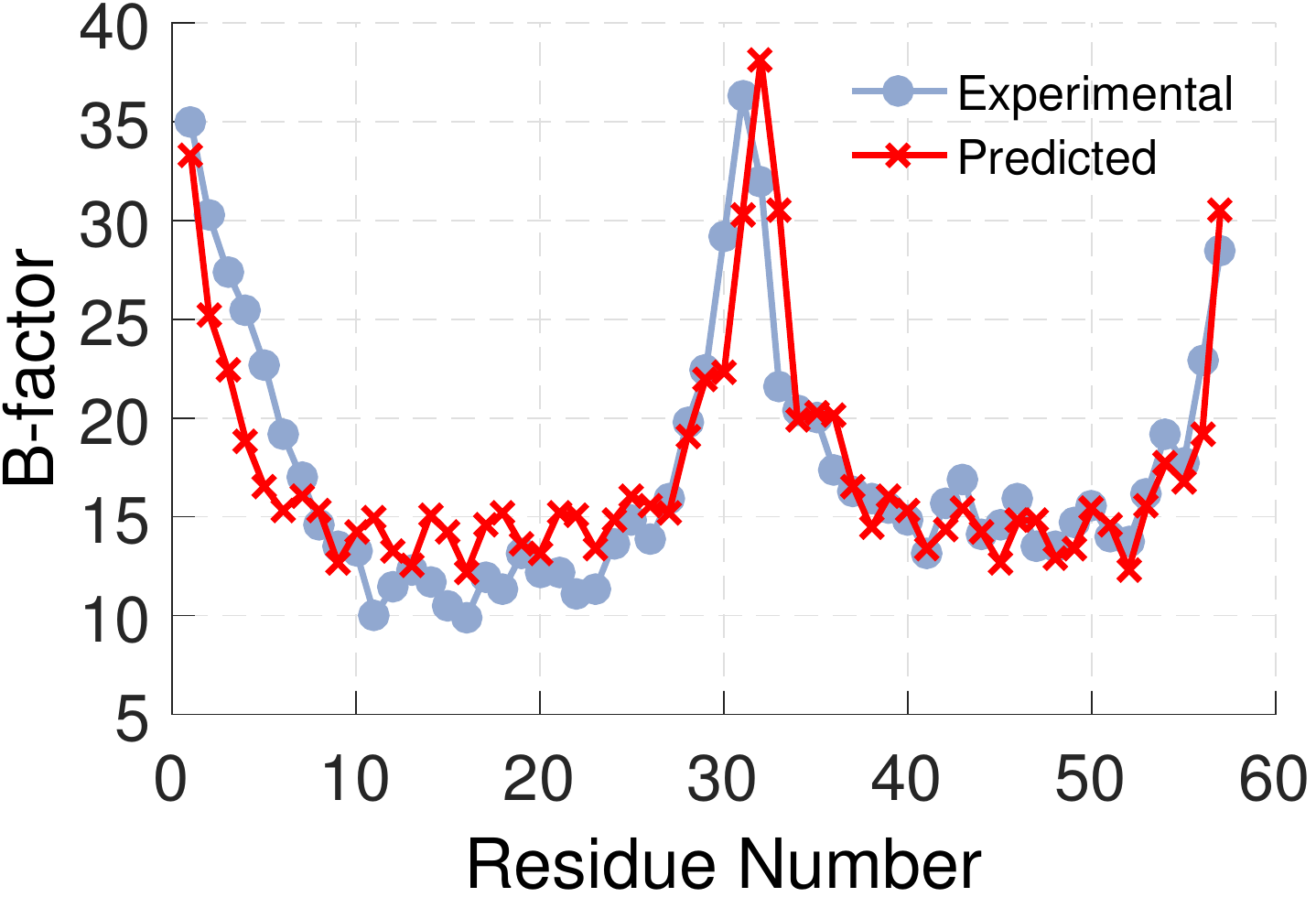}\quad
		\includegraphics[width=0.5\textwidth]{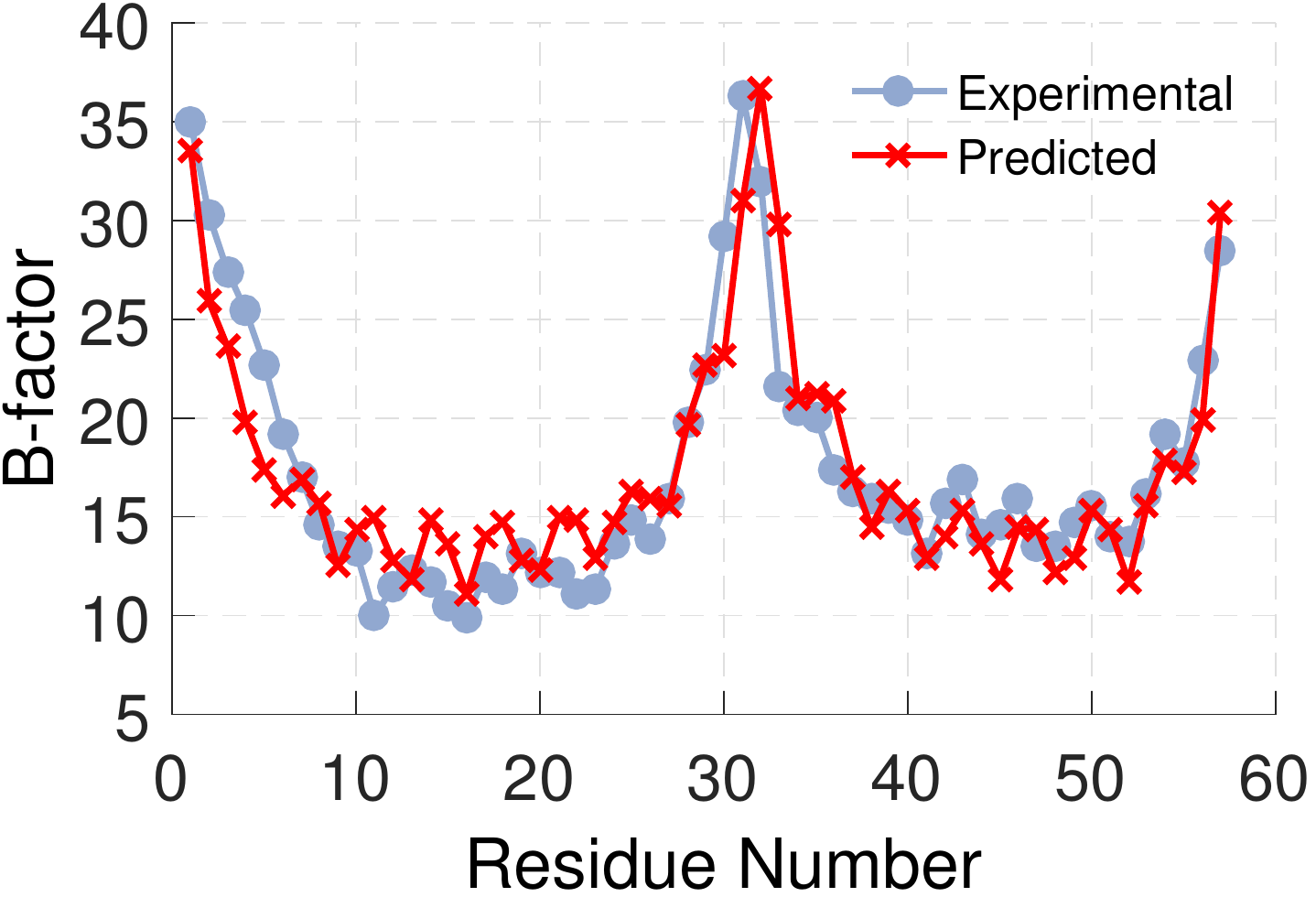}
		\caption{gFRI$^{22}$}
	\end{subfigure}
	\caption{Experimental B-factors (gray) vs predicted B-factors (red) of 1DF4 using the exponential (left) and Lorentz (right) correlation kernels. The optimal parameters for each type of B-factor prediction are described in Table \ref{tab.results}.
	}
	\label{fig.1df4}
\end{figure}

In the rest of this paper, we focus on the exploration of   gFRI  models associated $f^{\alpha\beta}_i$, i.e., $f^{11}_i$,  $f^{12}_i$,  $f^{21}_i$ and  $f^{22}_i$, and  denote these models as gFRI$^{\alpha\beta}$. We also study the performance  of gFRI for various kernel implementations, namely,  generalized exponential and  generalized Lorentz correlation kernels.

Due to the proportionality between the atomic flexibility index and the temperature factor at each atom, the theoretical B-factor at $i$th atom, $B_i^{\alpha\beta}$, can be expressed as a linear form
\begin{align}
B_i^{\alpha\beta}=a_{\alpha\beta} f^{\alpha\beta}_i+b_{\alpha\beta}, \quad \forall i=1,2,\dots,N; \alpha=1,2; \beta=1,2
\end{align}
where  constants $a_{\alpha\beta}$ and $b_{\alpha\beta}$ are independent of index $i$ and can be estimated by the following minimization process
\begin{align}\label{eq:miniming}
\min_{a_{\alpha\beta},b_{\alpha\beta}}\left\{\sum_{i=1}^N \left|B_i^{\alpha\beta}  - B_i^e\right|^2\right\},
\end{align}
where $B_i^e$ is the experimental B-factor for the $i$th   atom.
To quantitatively assess the performance of the proposed gFRI models for the B-factor prediction, we consider correlation
coefficient (CC)
\begin{align}
{\rm CC}=\frac{\sum_{i=1}^{N}\left(B_i^e-\hat{B}_i^e\right)\left(B_i^{\alpha\beta}-\hat{B}_i^{\alpha\beta}\right)}{\left[\sum_{i=1}^{N} \left(B_i^e-\hat{B}_i^e\right)^2
	\sum_{i=1}^{N} \left(B_i^{\alpha\beta}-\hat{B}_i^{\alpha\beta}\right)^2\right]^{1/2}},
\end{align}
where $\hat{B}^{\alpha\beta}$ and $\hat{B}^e$ are, respectively, the statistical averages of theoretical and experimental B-factors.

We consider a set of 364 proteins used in our earlier work \cite{KLXia:2015f} and coarse-grained C$_\alpha$ atoms in each protein. Therefore, we set $w_j=1$ and use a uniform  characteristic distance $\eta_j=\eta$ in all of our computations.

We firstly analyze the best parameter set for the B-factor prediction of each rigidity and flexibility type over a range of parameters.
{\color{black}Table \ref{tab.results} reveals the optimal parameters and the best MCCs for gFRI$^{\alpha\beta}$, with $\alpha=1,2$ and $\beta=1,2$.} For the sake of visualization, Fig. \ref{fig.cc} plots behavior of parameters for exponential and Lorentz kernels in B-factor predictions. It can be seen from Fig. \ref{fig.cc} that gFRI$^{21}$ and gFRI$^{22}$ models  are more sensitive to parameter $\eta$ than their gFRI$^{1\beta}$ counterparts. Despite having a fewer choices of fitting parameters, gFRI$^{21}$ and gFRI$^{22}$ models are still able to deliver B-factor predictions as  accuracy as those of gFRI$^{1\beta}$ models.

To further demonstrate the accuracy of each type of B-factor predictions for different correlations kernels, we plot predicted B-factors against the experimental ones for protein 1DF4 in Fig. \ref{fig.1df4}. In general, all B-factor prediction approaches produce a similar accuracy, especially when a Lorentz kernel is employed. Moreover, the utilization of exponential type of functions for gFRI$^{12}$  and gFRI$^{21}$  B-factor prediction types likely performs a little bit worse than the rest.

For an extended comparison, Table \ref{tab.results} also lists B-factor prediction performances of our earlier FRI \cite{Opron:2014} method, generalized GNM (gGNM)  \cite{KLXia:2015f}, generalized ANM (gANM)  \cite{KLXia:2015f} and the classic GNM \cite{Bahar:1997,Bahar:1998} approaches. The earlier FRI algorithm is the same as gFRI$^{11}$  in the present work  while omits the normalization process \eqref{normalization}.     By employing the same correlation kernel parameters as of  gFRI$^{11}$, the earlier FRI method gives  B-factor predictions similar to those of  gFRI$^{11}$. Specifically, MCCs produced by the previous FRI algorithm for exponential kernel and Lorentz kernel, are, respectively, 0.623 and 0.626. It is noted that the earlier FRI predicted B-factors over 365 proteins \cite{Opron:2014} while current methods employ the same data set with one left out, 1AGN, due to the unrealistic experimental B-factors.

The gGNM method \cite{KLXia:2015f} is an FRI kernel generalization of   GNM  \cite{Bahar:1997,Bahar:1998}. In this approach , the $i$th B-factor of a biomolecule can be defined as \cite{Bahar:1997,Bahar:1998}
\begin{align}
	B_i^{\text{gGNM}}=a_{\text{gGNM}}\left(\Gamma^{-1}\right)_{ii},\quad \forall i=1,2,\dots,N,
\end{align}
where $a_{\text{gGNM}}$ is a fitting parameter and $\left(\Gamma^{-1}\right)_{ii}$ is the $i$th diagonal element of the matrix inverse of the generalized Kirchhoff matrix  \cite{KLXia:2015f}
\begin{align}
	\Gamma_{ij}(\Phi)=\left\{
	\begin{array}{lr}
	-\Phi\left(\|\mathbf{r}_i-\mathbf{r}_j\|;\eta_{j}\right), & i\neq j,\vspace*{0.1in}\\
	-\sum_{j,j\neq i}^{N} 	\Gamma_{ij}(\Phi), & i=j.
	\end{array}
	\right.
\end{align}

Similarly, the gANM method \cite{KLXia:2015f} is an FRI kernel generalization of the classic ANM method  \cite{Atilgan:2001}. In this approach,
the generalized local $3\times 3$ Hessian matrix $H_{ij}$ is written as
\begin{align}
	H_{ij}=-\frac{\Phi\left(\|\mathbf{r}_i-\mathbf{r}_j\|;\eta_{ij}\right)}{\|\mathbf{r}_i-\mathbf{r}_j\|^2}\left[
	\begin{array}{ccc}
	(x_j-x_i)(x_j-x_i) & (x_j-x_i)(y_j-y_i) & (x_j-x_i)(z_j-z_i) \\
	(y_j-y_i)(x_j-x_i) & (y_j-y_i)(y_j-y_i) & (y_j-y_i)(z_j-z_i) \\
	(z_j-z_i)(x_j-x_i) & (z_j-z_i)(y_j-y_i) & (z_j-z_i)(z_j-z_i)
	\end{array}
	\right], \forall i\neq j.
\end{align}
We define the diagonal parts as $H_{ii}=-\sum_{i\neq j} H_{ij}, \forall i=1,2,\dots,N$. Therefore, the B-factor of $i$th $C_\alpha$ in a biomolecule is expressed as
\begin{align}
B_i^{\text{gANM}}=a_{\text{gANM}}\left(H^{-1}\right)_{ii}, \quad \forall i=1,2,\dots,N,
\end{align}
where $a_{\text{gANM}}$ is a fitting parameter and $(H^{-1})_{ii}$ is the $i$th diagonal element of the matrix inversion of a matrix formed by generalized local Hessian matrices.

As shown in Table \ref{tab.results}, the gGNM prediction with exponential kernel offers the MCC of 0.608  \cite{KLXia:2015f}  which is as good as gFRI$^{12}$ and gFRI$^{21}$ methods and is a bit worse than  gFRI$^{11}$ and gFRI$^{22}$ approaches. We still find a similar behavior when employing gGNM method with Lorentz kernels. In particular, its MCC is 0.622. On the other hand,  the gANM scheme has the worst performance among the considering prediction types. With an MCC found to be 0.518, the gANM approach with exponential kernel is much far behind on the accuracy. Since the gANM prediction with Lorentz type of functions is nowhere close to the acceptable level in term of the accuracy, the result for that case was not reported \cite{KLXia:2015f}. Finally, with a cutoff distance of 7\AA, the prediction of classic GNM method delivers an MCC of 0.565, which is better than that obtained by gANM, but not as good as any other methods in comparison. Since, GNM is one of the most popular and the most accurate methods for B-factor predictions \cite{LWYang:2008}, the current comparison shown in Table \ref{tab.results} indicates that FRI and gFRI  are  a new generation of  more accurate and robust methods for protein B-factor prediction.
\begin{figure}[!tb]
	\centering
	\includegraphics[width=0.3\columnwidth]{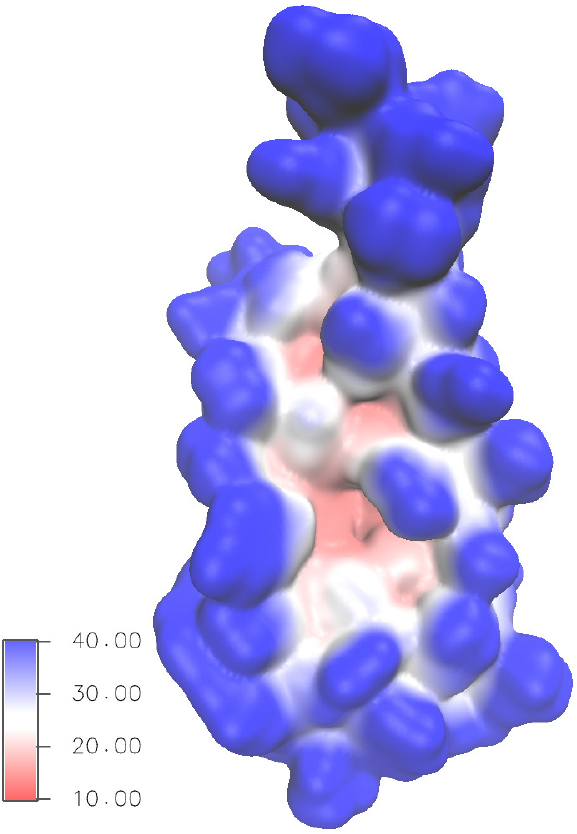}\quad \includegraphics[width=0.3\columnwidth]{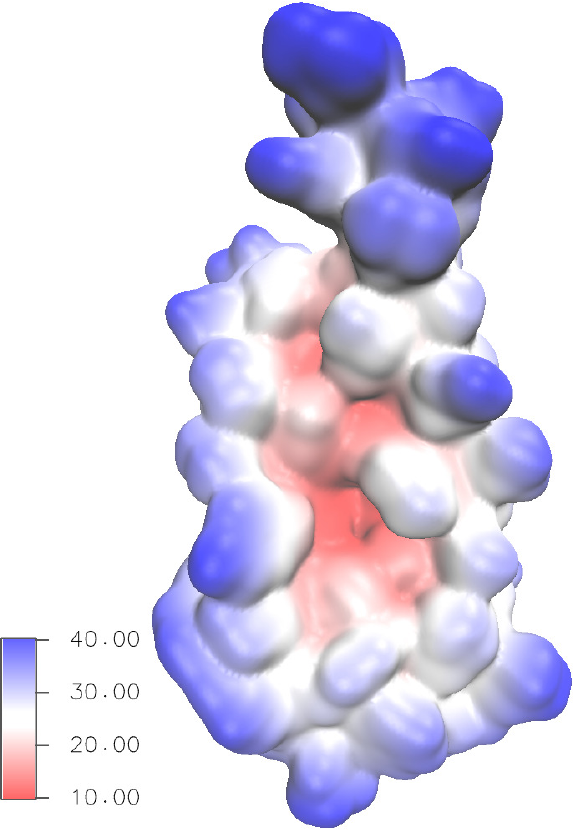}
	\caption{Molecular surface of 1DF4 colored by flexibility function with exponential kernel. Left: gFRI$^{11}$ with $\kappa=1.0$ and $\eta=3.0\text{~\AA}$. Right: gFRI$^{12}$  with $\kappa=1.0$ and $\eta=4.0\text{~\AA}$}
	\label{fig.surface}
\end{figure}

To further compare the B-factor performances between different methods, we introduce two types of continuous  flexibility functions.

\begin{align}\label{b1}
F^{\alpha1}(\mathbf{r})=\frac{a_{\alpha 1}} {\bar{\mu}^{\alpha}(\mathbf{r})} + b_{\alpha 1},
\end{align}
or
\begin{align}\label{b2}
F^{\alpha2}(\mathbf{r})=a_{\alpha 2}\left(1 - \bar{\mu}^{\alpha}(\mathbf{r})\right) + b_{\alpha 2},
\end{align}
where $a_{\alpha \beta}$  and $b_{\alpha \beta}$ are determined by minimization (\ref{eq:miniming}).  These flexibility functions are volumetric and can be projected onto a molecular surface for flexibility visualization. As mentioned earlier, rigidity densities provide excellent molecular surface representations and  one can  employ either rigidity density $\bar{\mu}^{1}(\mathbf{r})$ or  $\bar{\mu}^{2}(\mathbf{r})$ for surface generation. However, for the purpose of surface visualization,   rigidity density  $\bar{\mu}^{2}(\mathbf{r})$ is not suitable. The reason is that the rigidity formula \eqref{rigidity2} in the continuous form has to  avoid all the grid points near each atomic center.  Otherwise rigidity densities would be mostly 1 and the  corresponding flexibility functions would be mostly a constant on the molecular surface.  This hindrance can be remedied by using interpolation approach as discussed in our previous work \cite{KLXia:2013d}. In the present work, we only consider  $\bar{\mu}^{1}(\mathbf{r})$ for surface representation and employ $F^{11}(\mathbf{r})$ and $F^{12}(\mathbf{r})$ for flexibility visualization. The parameters for $F^{11}(\mathbf{r})$ and $F^{12}(\mathbf{r})$ follow those of  gFRI$^{11}$ and gFRI$^{12}$ as shown in Eqs. \eqref{b1} and \eqref{b2}, respectively.

  Figure  \ref{fig.surface} depicts the projection of flexibility functions $F^{1\beta}(\mathbf{r})$ onto the   isosurface  of rigidity density $\bar{\mu}^{1}(\mathbf{r})=0.05$  of protein 1DF4.  Parameters $\kappa=1$, $w_j=1$ and $\eta_j=0.5\text{~\AA}$ for all $j=1,\dots,N$ are used in $\bar{\mu}^{1}(\mathbf{r})$ for surface generation.
	  	Even though both gFRI$^{11}$ and gFRI$^{12}$ deliver similar B-factor predictions, with MCCs being 0.888 for gFRI$^{11}$ and is 0.889 for gFRI$^{12}$, their flexibility functions ($F^{1\beta}(\mathbf{r})$) behave  differently. It can be seen from Fig. \ref{fig.surface} that outer region of $F^{11}(\mathbf{r})$ projection contains higher values than its counterpart $F^{12}(\mathbf{r})$, while the inner region of both $F^{1\beta}(\mathbf{r})$ stays almost the same for both methods. This behavior is likely due to the fact that  $F^{11}(\mathbf{r})$ is constructed by  Eq. \eqref{b1}, which dramatically amplifies small rigidity densities far away from the center of mass of a molecule. In contrast, $F^{12}(\mathbf{r})$ is bound and well defined everywhere.

{\color{black} To predict the amplitudes and directions of atomic fluctuation, ANM \cite{Atilgan:2001} is commonly used. Another tool for such purpose is the anisotropic FRI (aFRI) proposed in our previous work \cite{Opron:2014}. It is interesting to know whether the present flexibility formulation \eqref{flexibility2} leads to  a new algorithm for protein anisotropic motion analysis.  To this end, we present a brief review of aFRI theory, which establishes notions for new formulation. In  ANM, the Hessian matrix is a global matrix containing $3N\times 3N$ elements with $N$ being a number of atoms. In our aFRI model, depending on one's interest, the size of the Hessian matrix can vary from $3\times 3$ for a completely local aFRI to $3N\times 3N$ for a completely global aFRI. To construct such a Hessian matrix, we partition all $N$ atoms in a molecule into a total of $M$ clusters $\{c_1,c_2,\dots,c_M\}$. Each cluster $c_k$ with $k=1,\dots,M$ has $N_k$ atoms so that $N=\sum_{k=1}^{M} N_k$. For convenience, we denote
	\begin{eqnarray}\label{eq:Anisorigidity1}
	\Phi^{ij}_{uv}  = \frac{\partial}{\partial u_i} \frac{\partial}{\partial v_j} \Phi( \|{\bf r}_i - {\bf  r}_j \|; \eta_{j} ), \quad  u,v= x, y, z; i,j =1,2,\cdots,N.
	\end{eqnarray}
	Note that  for each given $ij$, we define $\Phi^{ij}=\left( \Phi^{ij}_{uv} \right)$ as a local anisotropic matrix
	\begin{equation}
	\Phi^{ij}=\left(
	\begin{array}{ccc}
	\Phi^{ij}_{xx} & \Phi^{ij}_{xy}& \Phi^{ij}_{xz}\\
	\Phi^{ij}_{yx} & \Phi^{ij}_{yy}& \Phi^{ij}_{yz}\\
	\Phi^{ij}_{zx} & \Phi^{ij}_{zy}& \Phi^{ij}_{zz}
	\end{array}
	\right).
	\end{equation}

	In the anisotropic flexibility approach, a flexibility Hessian matrix ${\bf F}^{1}(c_k)$ for cluster $c_k$ is defined by
	\begin{eqnarray}\label{eq:Anisoflexibility}
	{\bf F}^{1}_{ij}(c_k)     =&  - \frac{1}{w_{j}} (\Phi^{ij})^{-1},                &\quad   i,j \in c_k; i\neq j;  u,v= x, y, z \\ \label{eq:Anisoflexibilityy3}
	{\bf F}^{1}_{ii}(c_k)=&   \sum_{j=1}^N \frac{1}{w_{j}} (\Phi^{ij})^{-1},  &\quad   i \in c_k;  u,v= x, y, z \\ \label{eq:Anisoflexibility4}
	{\bf F}^{1}_{ij}(c_k)=&  0,                                     &\quad   i,j \notin  c_k; u,v= x, y, z,
	\end{eqnarray}
	where $(\Phi^{ij})^{-1}$ denotes the unscaled inverse of matrix $\Phi^{ij}$ such that $\Phi^{ij}(\Phi^{ij})^{-1}=| \Phi^{ij}|$.

	Motivated by the new form of atomic flexibility indexes \eqref{flexibility2}, we propose another presentation for the flexibility Hessian matrix ${\bf F}^{2}(c_k)$ as follows
	\begin{eqnarray}\label{eq:Anisoflexibility5}
	{\bf F}^{2}_{ij}(c_k)     =&  - \frac{1}{w_{j}} |\Phi^{ij}|(J_{3} - \Phi^{ij}),                &\quad   i,j \in c_k; i\neq j;  u,v= x, y, z \\ 
	{\bf F}^{2}_{ii}(c_k)=&   \sum_{j=1}^N \frac{1}{w_{j}} |\Phi^{ij}|(J_3 - \Phi^{ij}),  &\quad   i \in c_k;  u,v= x, y, z \\ 
	{\bf F}^{2}_{ij}(c_k)=&  0,                                     &\quad   i,j \notin  c_k; u,v= x, y, z,
	\end{eqnarray}
	where $J_3$ is a $3\times3$ matrix with every element being  one.
	
	We can achieve $3N_k$ eigenvectors for $N_k$ atoms in cluster $c_k$ by diagonalizing ${\bf F}^{\alpha}(c_k)$, $\alpha=1,2$. Note that, the diagonal part ${\bf F}^{\alpha}_{ii}(c_k)$, $\alpha=1,2$, has inherent information of all atoms in the system. As a result, we can predict B-factors by employing Eq.  \eqref{eq:miniming} for a set of flexibility indexes collected from the diagonal parts
	\begin{eqnarray}\label{eq:Anisoflexibility2}
	f_i^{\rm AF_\alpha} &=&{\rm Tr} \left({\bf F}^{\alpha}(c_k)\right)^{ii},   \\
	&=&  \left({\bf F}^{\alpha}(c_k)\right)^{ii}_{xx}+ \left({\bf F}^{\alpha}(c_k)\right)^{ii}_{yy}+ \left({\bf F}^{\alpha}(c_k)\right)^{ii}_{zz}, \quad \alpha=1,2.
	\end{eqnarray}
	
	In this work, we   compare the protein anisotropic motion predictions by using completely global aFRI models based on the anisotropic flexibility associated with $f_i^{\rm AF_\alpha}$ and denote these models as $\rm aFRI^{\alpha}$, $\alpha=1,2$. Note that, model  $\rm aFRI^{1}$ is already discussed in our previous work \cite{Opron:2014}and used for a comparison. Figure \ref{fig.motions} depict the first three nontrivial isotropic modes  of aFRI$^{1}$, aFRI$^{2}$ and ANM for protein PBID: 2XHF. The Lorentz kernel is used for both aFRI algorithms with $w_j=1, \nu=2, $ and $ \eta_j=30\text{~\AA}, \forall j=1,\dots,N$ for aFRI$^{1}$, and $w_j=1, \nu=2, $and $ \eta_j=25\text{~\AA}, \forall j=1,\dots,N$ for aFRI$^{2}$. To obtain ANM prediction, we use Prody v1.8\cite{Bakan:2011} with default settings. It is interesting to see that each algorithm has its own set of collective protein motions. Since there is no exact answer to these fluctuation modes, we cannot conclude which motion prediction is right or wrong.
	
	\begin{figure}[!htb]
		\centering
		\begin{subfigure}[!htb]{0.8\textwidth}
			\centering
			\includegraphics[width=0.2\textwidth]{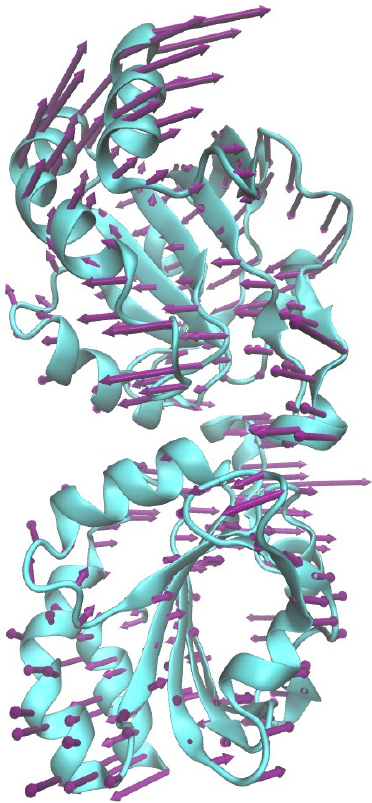}\quad
			\includegraphics[width=0.25\textwidth]{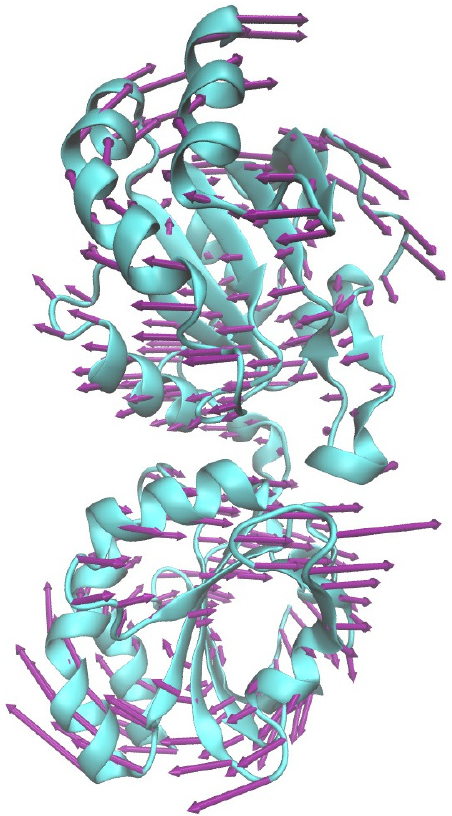}\quad
			\includegraphics[width=0.2\textwidth]{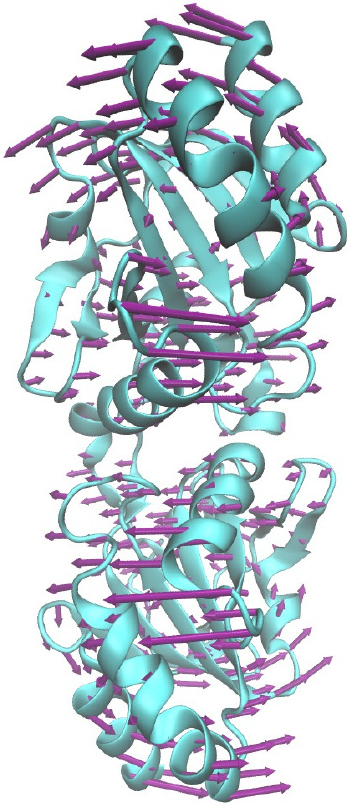}
			\caption{aFRI$^{1}$}
		\end{subfigure}
		\begin{subfigure}[!tb]{0.8\textwidth}
			\centering
			\includegraphics[width=0.2\textwidth]{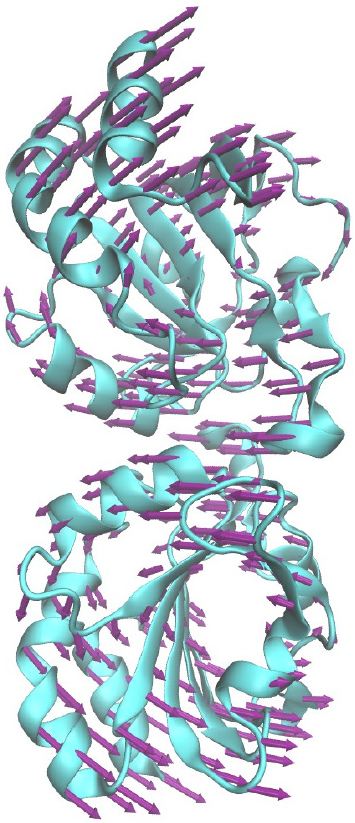}\quad
			\includegraphics[width=0.25\textwidth]{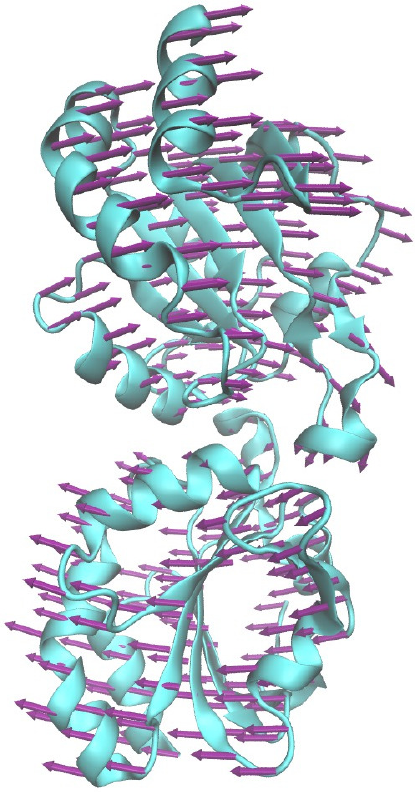}\quad
			\includegraphics[width=0.2\textwidth]{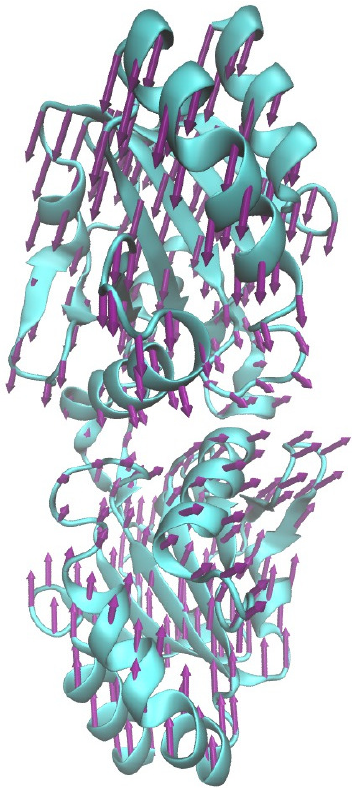}
			\caption{aFRI$^{2}$}
		\end{subfigure}
		\begin{subfigure}[!tb]{0.8\textwidth}
			\centering
			\includegraphics[width=0.2\textwidth]{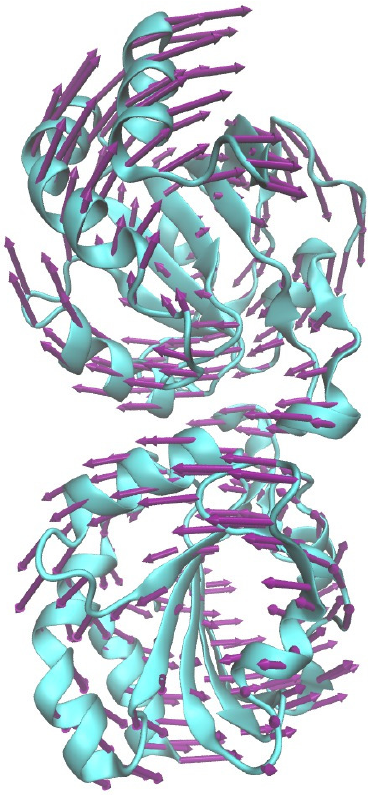}\quad
			\includegraphics[width=0.22\textwidth]{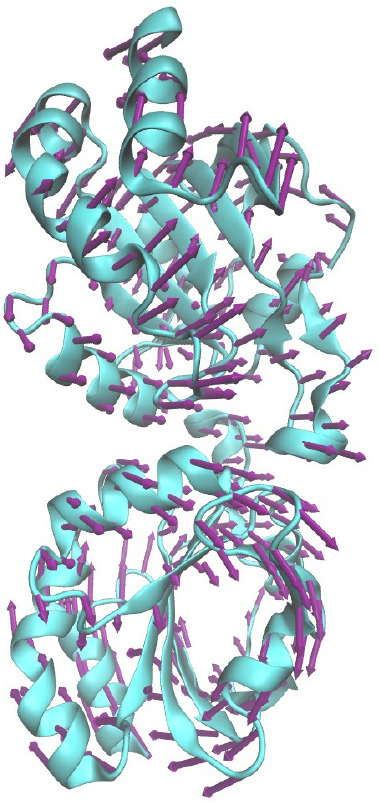}\quad
			\includegraphics[width=0.2\textwidth]{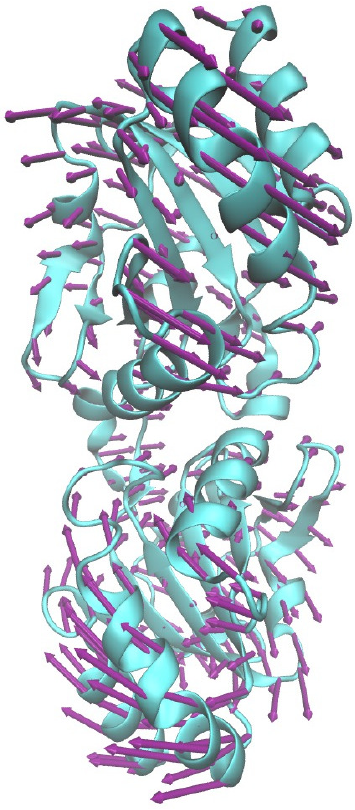}
			\caption{ANM}
		\end{subfigure}
		\caption{\color{black}Comparison of modes for protein PDB ID: 2XHF. The top row is generated by using the completely global aFRI$^{1}$ with $\nu=2$ and $\eta=30\text{~\AA~}$. The middle row is generated by using the completely global aFRI$^{2}$ with $\nu=2$ and $\eta=25\text{~\AA~}$. The bottom row is generated by using ANM with Prody v1.8\cite{Bakan:2011} using default settings.
		}
		\label{fig.motions}
	\end{figure}
	
}

It needs to point out that the proposed gFRI can be readily incorporated into our fFRI and mFRI methodologies.
Additionally, infinitely many possible atomic flexibility indexes  of a general functional form $f(\bar{\mu}^\alpha_i)$ can be designed.
For example, one can choose $f(\bar{\mu}^\alpha_i)= \Phi\left(\bar{\mu}^\alpha_i;\eta_{0}\right)$ with $\eta_0$ being a constant. However, it is not obvious how to design another distinct rigidity density formula.
A systematic analysis of these aspects is beyond the scope of the present work.

\section*{Acknowledgments}

This work was supported in part by NSF grant  IIS-1302285    and
MSU Center for Mathematical Molecular Biosciences Initiative.

%

\end{document}